\documentclass[12pt]{article}
\usepackage{epsf}
\usepackage{graphicx}

\catcode`\@=11
\@addtoreset{equation}{section}
\def\theequation{\thesection.\arabic{equation}}
\catcode`@=12
\relax

\def\F{{\cal F}}
\def\mathbb{\bf}

\begin{document}
\begin{titlepage}
\begin{flushright}
hep-th/0405128 \\
KEK-TH-954\\
May 2004
\end{flushright}
\vspace{0.5cm}
\begin{center}
{\Large \bf Gravitational Corrections for Supersymmetric Gauge \vskip 2mm 
Theories 
with Flavors via Matrix Models}
\\
\lineskip .75em
\vskip 1.5cm
{\large Hiroyuki Fuji \footnote[1]{\tt hfuji@post.kek.jp} and}  
{\large Shun'ya Mizoguchi \footnote[2]{\tt mizoguch@post.kek.jp}}\\  
\vskip 1.5em
{\large
    \vskip -1ex
       {\it Theory Division, Institute of Particle and Nuclear Studies} \\
       {\it High Energy Accelerator Research Organization (KEK)} \\
       {\it Tsukuba, Ibaraki 305-0801, Japan} \\}
\vskip 3.5em
\end{center}
\begin{abstract}
We study the gravitational corrections to the F-term in four-dimensional 
${\cal N}=1$ $U(N)$ gauge theories with flavors, 
using the Dijkgraaf-Vafa theory. We derive a compact formula for the 
annulus contribution in terms of the prime form on the matrix model curve. 
Remarkably, the full $\mbox{Tr}R\wedge R$ correction can be reproduced 
as a special momentum sector of a single $c=1$ CFT correlator, which 
closely resembles that in the bosonization of fermions on Riemann surfaces. 
The ${\cal N}=2$ limit of the torus contribution agrees with the 
multi-instanton calculations as well as the topological A-model result. 
The planar contributions, on the other hand, have no counterpart in 
the topological gauge theories, and we speculate about the origin 
of these terms.
\end{abstract}
\end{titlepage}

\baselineskip=0.6cm
\section{Introduction}
Since the discovery by Dijkgraaf and Vafa,  we have recognized that non-perturbative aspects of four-dimensional supersymmetric gauge 
theories can be studied via matrix models. 
In this framework the
effective superpotential for ${\cal N}=1$ supersymmetric 
gauge theories can be determined as the large $N$ free energy 
of a matrix model \cite{DV}, and by minimizing it
the non-perturbative vacua and their phase structures 
can be investigated
\cite{CSW0}.
This part of the proposal has been elegantly proven
by using the chiral ring relations of ${\cal N}=1$ supersymmetry  
in the generalized Konishi anomaly equations  
\cite{CDSW,KKM}.

On the other hand, the non-planar diagrams have been shown to
correspond to the gravitational corrections to the gauge theory
\cite{DST,KMT,OV,AC,INO,quiver},
in particular, the genus one free energy of the matrix model 
computes the gravitational F-term 
\begin{eqnarray}
\frac{1}{16\pi^2}\int d^4x \F_{1}(S) {\rm Tr}R\wedge R.
\end{eqnarray}
Recently, there has been much progress in
the analysis of ${\cal N}=2$ gravitational F-terms
in terms of the multi-instanton calculations \cite{Nekrasov}
and geometric engineering \cite{Iq-Kas,EK,Tachikawa}.
To compare them with the matrix model results, it is 
essential to know the precise form of the 
gravitational F-terms 
computed using the matrix model.
For the pure gauge theory case, it has been checked that 
the torus free energy $\F_1$ coincides with 
the topological partition function \cite{MW} of the Donaldson-Witten 
theory, and hence the ${\cal N}=2$ gravitational F-terms \cite{DST,KMT}.
The planar free energy is also known to contribute to the gravitational 
corrections \cite{DGOVZ}.

This framework can be extended to 
${\cal N}=1$ supersymmetric gauge theories with $N_f$ flavors 
in the fundamental representation
\cite{NSW,CSW,matter}.
The dual matrix model consists of a bosonic matrix and vectors 
which correspond to the adjoint scalar $\phi$ 
and $N_f$ flavor fields 
$q^{I}$, $\tilde{q}_I$
($I=1,\cdots ,N_f$) of the gauge theory, respectively.
In the vector coupled matrix model,
there are two kinds loops 
with $\hat{N}$ and $N_f$ indices;
the former are summed up 
but the latter remain as boundaries
of Feynman 
diagrams. 
It looks as if the model has not only the closed string 
but also the open string sectors \cite{Kazakov}.
In order to evaluate $R^2$ terms for this gauge theory,
it is necessary to consider both the torus and annulus 
contributions.

In this paper, we evaluate the $R^2$ correction terms for 
${\cal N}=1$ $U(N)$ gauge theories with $N_f$ flavors.
Using the ordinary large $N$ analysis \cite{BIPZ}, 
we derive a compact formula for the annulus contribution to 
the correction in terms of the prime form on the matrix model curve.
Remarkably, the full $\mbox{Tr}R\wedge R$ correction containing
the torus as well as all the planar contributions is reproduced as 
a special momentum sector of a single $c=1$ CFT correlator, which 
closely resembles that in the bosonization of fermions on Riemann 
surfaces.  It is in accord with the recent observation 
in topological string theory 
that the non-compact B-branes correspond to 
fermions in a chiral boson theory on a hyper-elliptic curve
\cite{integrable,Sakai}.

The plan of this paper is as follows:
In section 2 we briefly review the Dijkgraaf-Vafa theory 
including a string theory derivation of gravitational corrections. 
In section 3 we compute the planar gravitational corrections for the 
${\cal N}=1$ $U(N)$ gauge theories. In particular, we derive in section 3.4 
a compact formula for the annulus contributions in terms of the prime form 
on the matrix model curve. Section 4 is devoted to some examples,
in which our formula is confirmed explicitly.  
In section 5 we show that the full $\mbox{Tr}R\wedge R$ 
gravitational correction, including both the torus and planar contributions, 
is reproduced as a single chiral correlator of a $c=1$ conformal field 
theory, which closely resembles that in the bosonization of fermions 
on Riemann surfaces. In section 6 we consider the ${\cal N}=2$ limit,
and speculate about the origin of the planar contributions. 
Finally,  we conclude our paper with a summary and outlook 
on future work in section 7. The appendices contain some technical 
details which we need in the text.

\section{Gravitational F-terms from Matrix Models}
\subsection{The Dijkgraaf-Vafa Theory with Flavors}
The original proposal of Dijkgraaf and Vafa\cite{DV} was summarized 
as follows :
\begin{itemize}
\item[1.] The low energy effective superpotential of ${\cal N}=1$ gauge 
theories can be computed by summing over the planar diagrams of the 
matrix model with the same tree-level potential.
\item[2.] The non-planar diagrams compute the gravitational F-terms 
for these theories.
\end{itemize}
The first statement was proven in \cite{CDSW}, and was generalized to the cases 
with flavors in \cite{CSW}. The latter part was also supported by many arguments 
\cite{AC,INO}, and was explicitly confirmed in the pure gauge theory cases 
\cite{DST,KMT}.

Let us consider the ${\cal N}=1$ $U(N)$ gauge theory coupled to matter 
superfields $q_I, \tilde{q}_I$ ($I=1,\cdots, N_f$) in the 
(anti-)fundamental representation with the superpotential \cite{NSW}
\begin{eqnarray}
V(\phi,q_I)&=&
\mbox{tr}W(\phi)
+\sum_{I=1}^{N_f}\tilde{q}_I(\phi-m_I)q_I,\nonumber\\
W(\phi)&=&\sum_{p=1}^{n+1} \frac{g_p}{p} \phi^p.
\end{eqnarray}
$\phi$ is the adjoint chiral superfield.
In the classical vacuum the gauge group is broken as 
\begin{eqnarray}
U(N)&\rightarrow&\prod_{i=1}^n U(N_i),~~~~~~
\sum_{i=1}^n N_i=N.
\end{eqnarray}
The claim of \cite{DV} is that the non-perturbative vacuum 
structure of this theory can be analyzed by a vector coupled 
matrix model with action
\begin{eqnarray}
S_{\mbox{\scriptsize matrix}}(\Phi,Q_I,\tilde{Q}_I)
={\rm Tr}_{\hat{N}\times\hat{N}}W(\Phi)+
\sum_{I=1}^{N_f}\tilde{Q}_I(\Phi-m_I)Q_I
\end{eqnarray}
where $\Phi$ is an $\hat{N}\times\hat{N}$ hermitian matrix
and $Q_I$, $\tilde{Q}_I$ are $\hat{N}$-component vectors.
The vector-matrix coupling $\tilde{Q}_I\Phi Q_I$ leads to Feynman 
diagrams with boundaries in various topologies.
 \begin{figure}[ht]
  \centerline{\epsfxsize=15cm \epsfbox{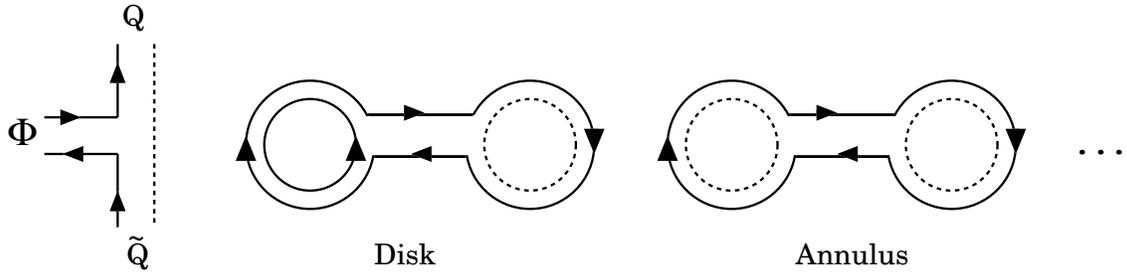}}
  \vskip 3mm
  \caption{Feynman diagrams with boundaries.}
  \label{Boundary interaction}
 \end{figure}

In the low energy effective theory,
the glueball superfields $S_i=\frac{1}{32\pi^2}{\rm tr}_{SU(N_i)}W_{\alpha}W^{\alpha}$ $(i=1,\ldots,n)$
play the role of the fundamental fields.
According to the proposal,
$S_i$ is identified with the 't Hooft coupling $g_s \hat{N}_i$ for each $i$, 
where $g_s$ is the matrix model coupling 
constant and $\hat{N}_i$ is the number of eigenvalues 
distributed on the $i$-th cut. 
Under this identification, 
the effective superpotential for 
the gauge theory is given by
\begin{eqnarray}
W_{\rm eff}(S_i)=
\sum_{i=1}^{n}\left(
N_i\frac{\partial \F_0^{(0)}(S_i)}{\partial S_i}+\tau S_i
\right)+\F_0^{(1)}(S_i)
\label{eff pot}
\end{eqnarray}
where $\F_0^{(0)}(S_i)$ and $\F_0^{(1)}(S_i)$ are given by the large $\hat{N}$
sphere and disk free energies of the matrix model, respectively. 
$\tau$ is the bare coupling.\footnote{The definition of $W_{\rm eff}(S_i)$
(\ref{eff pot}) has an ambiguity in the  linear terms in $S_i$ depending on
the choices of the integration constants in the free energy (See e.g. \cite{CSW}.).}   
By extremizing this superpotential, one can analyze 
the non-perturbative vacuum structure of the gauge theory 
\cite{CSW0}.

\subsection{String Theory Derivation of Gravitational F-terms}
To extend the analysis to the study of the gravitational F-terms, 
it is useful to consider how they arise in string theory. 
The gauge theory setup above can be realized in 
type IIB string theory on a Calabi-Yau three-fold 
\cite{CIV,CV}
\begin{eqnarray}
W^{\prime}(z)^2+y^2+v^2+w^2=0,
\quad 
(z,y,v,w)\in \mathbb{C}^4
\end{eqnarray}
with 
$n$ singular points. 
By blowing them up, we obtain a smooth geometry with 
exceptional 2-cycles $C_i$ ($i=1,\cdots,n$).
We then consider
$N$ D5-branes wrapped around 
$C_i$'s and $N_f$ D5-branes around non-compact 2-cycles 
$C_I$ ($I=1,\cdots,N_f$).
The ${\cal N}=1$ gauge theory above is 
realized on the space-time filling world-volume of the D5-branes.
 \begin{figure}[ht]
  \centerline{\epsfxsize=15cm \epsfbox{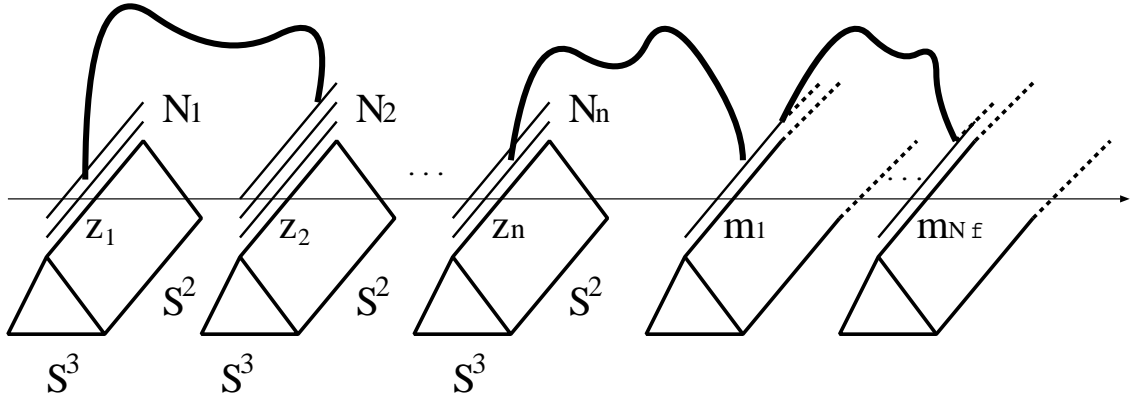}}
  \vskip 3mm
  \caption{The brane configuration in the resolved geometry.}
  \label{resolved geometry}
 \end{figure}

To  evaluate the F-term corrections
it is advantageous to utilize the hybrid formalism \cite{Berkovits,OV}. 
The string Lagrangian density for the four-dimensional part is 
given by
\begin{eqnarray}
{\cal L}=\frac{1}{2}\partial X^{\mu}\bar{\partial}X_{\mu}
+p_{\alpha}\bar{\partial}\theta^{\alpha}
+p_{\dot{\alpha}}\bar{\partial}\theta^{\dot{\alpha}}
+\bar{p}_{\alpha}\bar{\partial}\bar{\theta}^{\alpha}
+\bar{p}_{\dot{\alpha}}\bar{\partial}\bar{\theta}^{\dot{\alpha}},
\end{eqnarray}
where $p$, $\bar{p}$, $\theta$ and $\bar{\theta}$ are the fermionic fields.
Inserting two gluino vertex operators 
$\oint_{\gamma} W^{\alpha}p_{\alpha}$
on the boundaries $\gamma$ of the world-sheet, 
the stringy corrections lead to the F-terms 
containing  the glueball superfields.

The string world-sheet can end either on $N$ compact  
or $N_f$ non-compact branes in the Calabi-Yau direction. 
Resorting to the fermion zero-mode arguments \cite{AGNT,BCOV}, 
we can conclude that 
only the planar string world-sheets with at most one boundary 
ending on the non-compact branes can contribute to the
effective superpotential \cite{DV}. A simple combinatorial argument then yields 
the effective superpotential (\ref{eff pot}) where 
the sphere and disk free energies of string theory are 
identified with those of the matrix model.

\begin{figure}[ht]
  \centerline{\epsfxsize=7cm \epsfbox{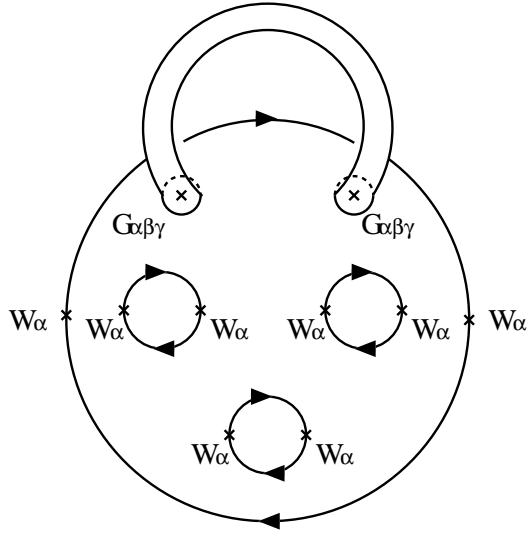}}
  \vskip 3mm
  \caption{Gravitational corrections from the torus.}
  \label{Torus}
 \end{figure}

\begin{figure}[ht]
  \centerline{\epsfxsize=17cm \epsfbox{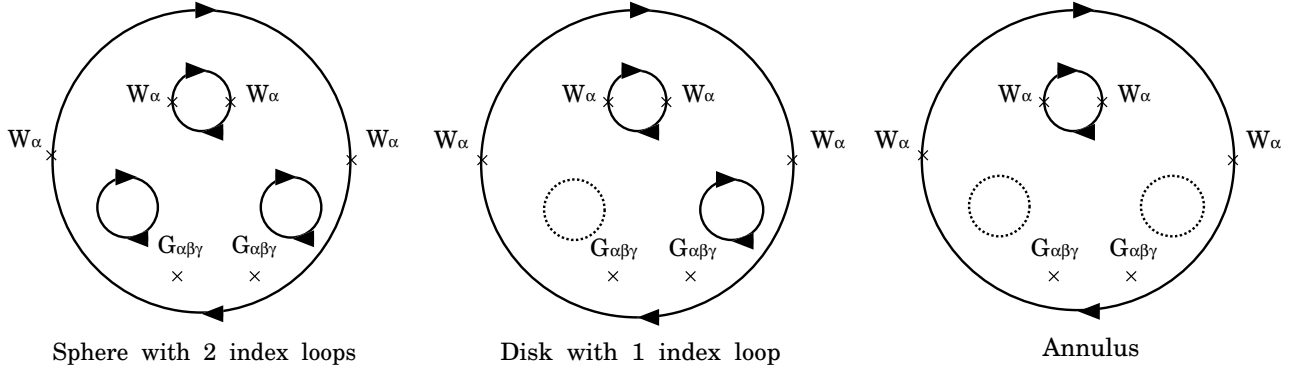}}
  \vskip 3mm
  \caption{Three types of planar string world-sheets.}
  \label{3Annular}
 \end{figure}

The gravitational F-term corrections are given by 
inserting RR vertex operators on the bulk of the string world-sheet.
A candidate for such an operator 
is the self-dual gravitino vertex operator \cite{OV,AC}
\begin{eqnarray}
\int  G^{\alpha\beta\gamma}\biggl(
p_{\alpha}X_{\beta\dot{\beta}}\bar{\partial}X_{\gamma\dot{\gamma}}
+\bar{p}_{\alpha}X_{\beta\dot{\beta}}\partial X_{\gamma\dot{\gamma}}
\biggr)
\epsilon^{\dot{\beta}\dot{\gamma}}
+\int G^{\alpha\beta\gamma}p_{\alpha}\bar{p}_{\beta}
(\theta_{\gamma}-\bar{\theta}_{\gamma}).
\end{eqnarray}
Due to the chiral ring relations 
\begin{eqnarray}
\{
W_{\alpha},W_{\beta}
\}\equiv 
2G_{\alpha\beta\gamma}W^{\gamma},
\quad
G^4\equiv 0
\quad ({\rm mod}\;\bar{D}),
\end{eqnarray}
at most two $G_{\alpha\beta\gamma}$ insertions can contribute to
the gravitational F-term corrections.
Such $G^2$ terms are given by 
the torus (Figure \ref{Torus}) as well as the planar world-sheet corrections 
(Figure \ref{3Annular}).
In particular, there are three types of planar world-sheets depending on 
which (compact or non-compact) branes the string ends on.


The gravitational F-term containing $G^2$ takes the form 
\begin{eqnarray}
\int d^4xd^2\theta \F_{\chi=0}(S_i)G^2,
\end{eqnarray}
where $\F_{\chi=0}(S_i)$ ($=$ the planar $+$ torus gravitational 
corrections)
is the contribution  
from the $\chi=0$ string world-sheets.\footnote{Here only the empty loops 
are understood as boundaries in the definition of the Euler number $\chi$.}
If we further consider a self-dual graviphoton background 
$F_{\alpha\beta}$,
then we also have gravitational F-terms from the 
graviphoton fields.
Such F-terms are obtained by decomposing ${\cal N}=2$
gravitational F-terms into the ${\cal N}=1$ F-terms.
It was shown in \cite{OV} that the superstring computation of 
these Feynman diagrams is identical to that in the field theory 
limit, and the combinatorial factors can be calculated 
in the associated matrix model.

In the following sections, we will derive a formula 
for the free energy $\F_{\chi=0}(S_i)$.

\section{Planar Gravitational Corrections}
\subsection{Gravitational Corrections from Sphere and Disk Diagrams}
There are three kinds of planar diagrams which 
contribute to the $\mbox{Tr}R\wedge R$ correction (Figure \ref{3Annular}).
The first and second types of contributions are determined  
by the sphere and the disk free energies through 
the following combinatorial arguments:
For the unbroken gauge group case (one-cut case), 
the coefficient of planar gravitational corrections yields
\begin{eqnarray}
\sum_{h=1}^\infty
\phantom{}_h C_2 \F_{0,h}^{(0)} 
(\mbox{tr}_{SU(N)}1)^2 S^{h-2}=
\frac{N^2}{2}\frac{\partial^2 {\cal F}_0^{(0)}(S)}{\partial S^2},
\quad {\cal F}_0^{(0)}(S)=\sum_{h=1}^{\infty}\F_{0,h}^{(0)} S^h
\end{eqnarray}
for the first type, and 
\begin{eqnarray}
\sum_{h=1}^\infty
\phantom{}_h C_1 \F_{h}^{(0)} 
(\mbox{tr}_{SU(N)}1)^2 S^{h-1}
=N\frac{\partial {\cal F}_0^{(1)}(S)}{\partial S}, 
\quad {\cal F}_0^{(1)}(S)=\sum_{h=1}^{\infty}\F_{0,h}^{(1)} S^h
\end{eqnarray}
for the second type.
$\F_{0,h}^{(0)}$ and $\F_{0,h}^{(1)}$
are the symmetric factors of Feynman diagrams
with $h$ boundaries
for the sphere and disk topologies, respectively.
Generalizing to the cases of arbitrary breaking patterns,
we have 
\begin{eqnarray}
&&
\frac{1}{2}\sum_{i,j=1}^n N_i N_j \frac{\partial^2 
\F^{(0)}_0
}{\partial S_i \partial S_j}
\quad (
\mbox{for a sphere with 2 index loops}
),
\\
&& 
\sum_{i=1}^n N_i \frac{\partial 
\F^{(1)}_0
}{\partial S_i}
\quad\quad\quad\quad\; (
\mbox{for a disk with 1 index loop}
).
\end{eqnarray}

The third type of gravitational corrections come from annulus 
diagrams, which we will compute below using the familiar 
large $N$ technique in matrix models \cite{BIPZ}.

\subsection{Planar Diagrams with Boundaries}
Let $ \Phi $ be an $\hat N\times\hat N$ hermitian
matrix, and $Q^I$ $(I=1,\ldots,N_f)$ be 
$\hat N$-dimensional complex vectors. 
We consider the partition function 
\begin{eqnarray}Z&=&\frac1{\mbox{vol}~\!U(\hat N)}\int d \Phi  \prod_{I=1}^{N_f}(d Q_I d\tilde Q_I) 
\nonumber\\
&&\times\exp \left[-\frac1{g_s}
\left(
{\rm Tr}W( \Phi ) + 
\sum_{I=1}^{N_f} (\tilde Q_I  \Phi  Q^I - m_I \tilde Q_I Q^I)\right)\right]
\end{eqnarray}
with a polynomial tree level potential 
$W(\Phi)$.
It allows  a topological expansion \cite{tHooft}
\begin{eqnarray}
Z&=&\exp(g_s^{-2}\F),\nonumber\\
\F&=&\sum_{g=0}^{\infty}\sum_{k=0}^{\infty}g_s^{2g+k} \F_{g}^{(k)},
\end{eqnarray}
where $\F_{g}^{(k)}$ is the contributions from the connected genus $g$
diagrams with $k$ boundaries. $\F_0^{(0)}$ also includes the 
non-perturbative piece coming from the gauge volume \cite{OVlargeN}.

To evaluate the annulus amplitude $\F_0^{(2)}$, we carry out the 
$d Q_I d\tilde Q_I$ integrations first. This yields 
\begin{eqnarray}
Z&\propto&\int d \Phi  
\exp \left[-\frac1{g_s} {\rm Tr}\left(W( \Phi ) + g_s 
\sum_{I=1}^{N_f} \log ( \Phi  - m_I)
\right)
\right]
\end{eqnarray}
up to an irrelevant multiplicative constant which is independent of $ \Phi $, 
$W( \Phi )$ or $m_I$'s. This integrations organize a resummation of diagrams. 
The $g_s$ factor of the $\log$ potential indicates that there 
is a matter ($=Q\tilde Q$ ) loop at each $\log$ vertex. Thus, instead of 
$Z$,  we consider the following double expansion 
\begin{eqnarray}
Z(g_s,\epsilon)&:=&\int d \Phi  
\exp \left[-\frac1{g_s} {\rm Tr}\left(W( \Phi ) + \epsilon \sum_{I=1}^{N_f} \log ( \Phi  - m_I)
\right)
\right]\label{Zgsepsilon}\\
&:=&\exp (g_s^{-2}\F(g_s,\epsilon)), \nonumber\\
\F(g_s,\epsilon)&=&\sum_{g=0}^\infty g_s^{2g}\F_{g}(\epsilon),\\
\F_{g}(\epsilon)&=&\sum_{k=0}^\infty \epsilon^k \F_{g}^{(k)} 
\end{eqnarray}
and set $\epsilon = g_s$. Then $\F_1^{(0)}$ and $\F_0^{(2)}$ are of same $g_s^2$ order
in the ordinary $g_s$ expansion of the free energy; the former is 
the torus, whereas the latter is the annulus amplitude.  

One of the aims of this paper is to determine the precise form of 
$\F_0^{(2)}$ for general $n$ and for a general polynomial potential $W(z)$,
in terms of the language of Riemann surfaces that the matrix model 
defines.  $\F_0^{(k)}$ $(k=1,2,\ldots)$ can be extracted from $Z(g_s, \epsilon)$ 
as follows: Writing the expectation value of any function $f(\Phi)$ of $\Phi$ 
with respect to the measure (\ref{Zgsepsilon}) as 
$\left<
f(\Phi)
\right>_{g_s,\epsilon}$,
we have
\begin{eqnarray}\F_0^{(k)}&=&
\frac1{k!}\left.\frac{\partial^k \F_0(\epsilon)}{\partial \epsilon^k}
\right|_{\epsilon=0}\nonumber\\
&=&
\frac1{k!}\left.
\lim_{g_s\rightarrow 0}
\frac{\partial^k \F(g_s,\epsilon)}{\partial \epsilon^k}
\right|_{\epsilon=0}\nonumber\\
&=&
\frac1{k!}\left.
\lim_{g_s\rightarrow 0}\frac{\partial^{k-1}}{\partial\epsilon^{k-1}}\left(
-g_s\left<
\mbox{Tr}{\textstyle \sum_{I=1}^{N_f}}\log(\Phi-m_I)
\right>_{g_s,\epsilon}
\right)
\right|_{\epsilon=0}.\end{eqnarray}
Keeping 
$S:= g_s \hat N$ fixed, $g_s\rightarrow 0$ implies $\hat N\rightarrow
\infty$.  In this limit the expectation value can be written as 
\begin{eqnarray}\lim_{g_s\rightarrow 0}\frac1{\hat N}
\left<\mbox{Tr}  f(\Phi)\right>_{g_s,\epsilon}
&=&
\int_{-\infty}^\infty d\lambda \rho(\lambda,\epsilon)f(\lambda) \end{eqnarray}
for any $f(\Phi)$, where $\rho(\lambda,\epsilon)$ is the eigenvalue density function 
of $\Phi$ normalized as 
\begin{eqnarray}
\int_{-\infty}^\infty d\lambda \rho(\lambda,\epsilon)=1.
\label{normalization}
\end{eqnarray} 
Therefore, expanding $\rho(\lambda,\epsilon)$ as
\begin{eqnarray} 
\rho(\lambda,\epsilon)&=&\sum_{k=0}^\infty \epsilon^k \rho^{(k)}(\lambda),
\end{eqnarray}we find 
\begin{eqnarray}\F_0^{(k)}&=&\left.
\frac1{k!}\frac{\partial^{k-1}}{\partial\epsilon^{k-1}}
\left(-S\int_{-\infty}^\infty d\lambda \rho(\lambda,\epsilon)
 \sum_{I=1}^{N_f}\log(\lambda-m_I)\right)\right|_{\epsilon=0}
 \nonumber\\
 &=&
-\frac Sk
\int_{-\infty}^\infty d\lambda \rho^{(k-1)}(\lambda)
 \sum_{I=1}^{N_f}\log(\lambda-m_I),
 \label{F0kintegral}
\end{eqnarray}
in particular, the disk \cite{NSW,CSW}
\begin{eqnarray}\F_0^{(1)}&=&-S
\int_{-\infty}^\infty d\lambda \rho^{(0)}(\lambda)
 \sum_{I=1}^{N_f}\log(\lambda-m_I).
 \label{F01integral}
\end{eqnarray}
and the annulus  
\begin{eqnarray}\F_0^{(2)}&=&-\frac S2
\int_{-\infty}^\infty d\lambda \rho^{(1)}(\lambda)
 \sum_{I=1}^{N_f}\log(\lambda-m_I).
 \label{F02integral}
\end{eqnarray}$\rho^{(k-1)}(\lambda)$ can be computed by means of the standard large $N$ 
technique for evaluating the planar diagrams \cite{BIPZ}.  
 
To be rigorous, one would need to show that the $g_s\rightarrow 0$ limit and 
the operation of $\frac\partial{\partial\epsilon}$ commute. In Appendix A we 
will give a different derivation of the formula (\ref{F02integral}) for an
arbitrary polynomial potential $W(\Phi)$ using Riemann's bilinear identity,
thereby providing an alternative proof of it.

\subsection{The Large $N$ Technique}
\label{largeN}
We will now briefly review the large $N$ ($=\hat N$) 
technique developed in  \cite{BIPZ}
to fix our notation.  (See also \cite{ACKM}.) 
As we introduced above, $\rho(\lambda,g_s)$ is the 
density function of the eigenvalue $\lambda$ of $\Phi$ normalized as 
(\ref{normalization}). 
The nonzero support of $\rho(\lambda,g_s)$ 
(as a function of $\lambda$) consists of several disjoint intervals on the 
real $\lambda$ axis. These intervals are the branch cuts of 
the {\it resolvent} $\omega(z,g_s)$ defined by
\begin{eqnarray}\omega(z,g_s):= \int_{-\infty}^{\infty} 
d\lambda \frac{\rho(\lambda,g_s)}{z-\lambda}
\label{omegadef}\end{eqnarray}
on the complex $z$ plane. 
(\ref{omegadef}) is equivalent to 
\begin{eqnarray}
\omega(\lambda+i0,g_s)-\omega(\lambda-i0,g_s)&=&-2\pi i \rho(\lambda,g_s)
\end{eqnarray}
in the boundary value representation. 
If the number of cuts is $n$, $\omega(z,g_s)dz$
is a meromorphic differential on a hyper-elliptic curve
of genus $n-1$ with a parameter $g_s$.

Let ${{\it \Sigma}^{g_s}_{n-1}}$ be a family of genus $n-1$ hyper-elliptic 
curves defined by the equation
\begin{eqnarray}
y^2=\prod_{i=1}^{2n}(z-a_i(g_s))~~~
~(i=1,\ldots,2n)
\label{y}
\end{eqnarray}
with a parameter $g_s$. We denote
by $\sqrt{\prod_{i=1}^{2n}(z-a_i(g_s))}$  a branch of $y$ such that 
$y>0$ when $z\in\mathbb{R}$, $z\rightarrow\infty$. 
We call it the first sheet, 
and the other the second sheet.

\begin{figure}
\centering
\resizebox{0.65\textwidth}{!}{%
  \rotatebox{0}{\includegraphics{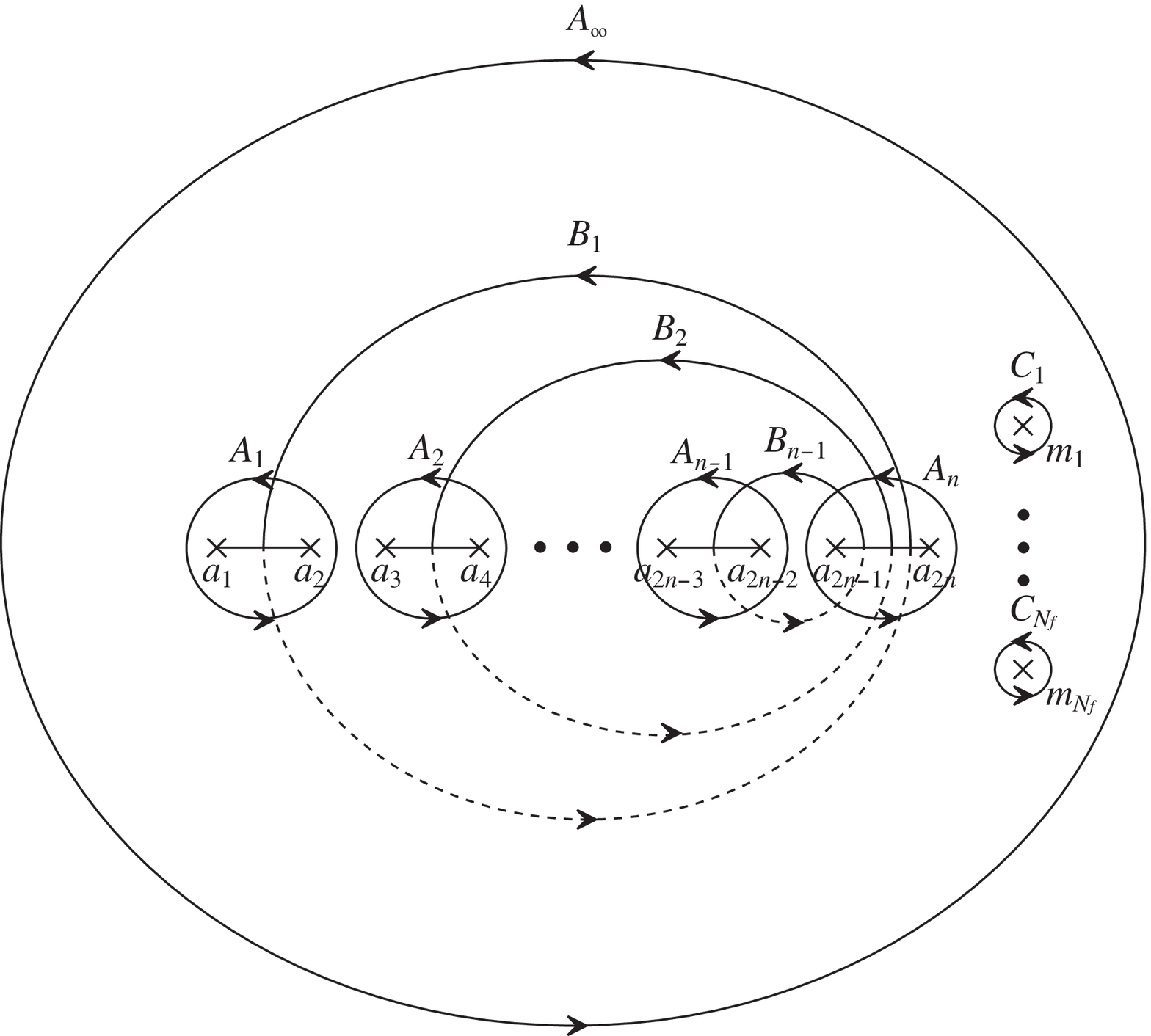}}}\\
\vskip 3mm
\resizebox{0.65\textwidth}{!}{%
  \rotatebox{0}{\includegraphics{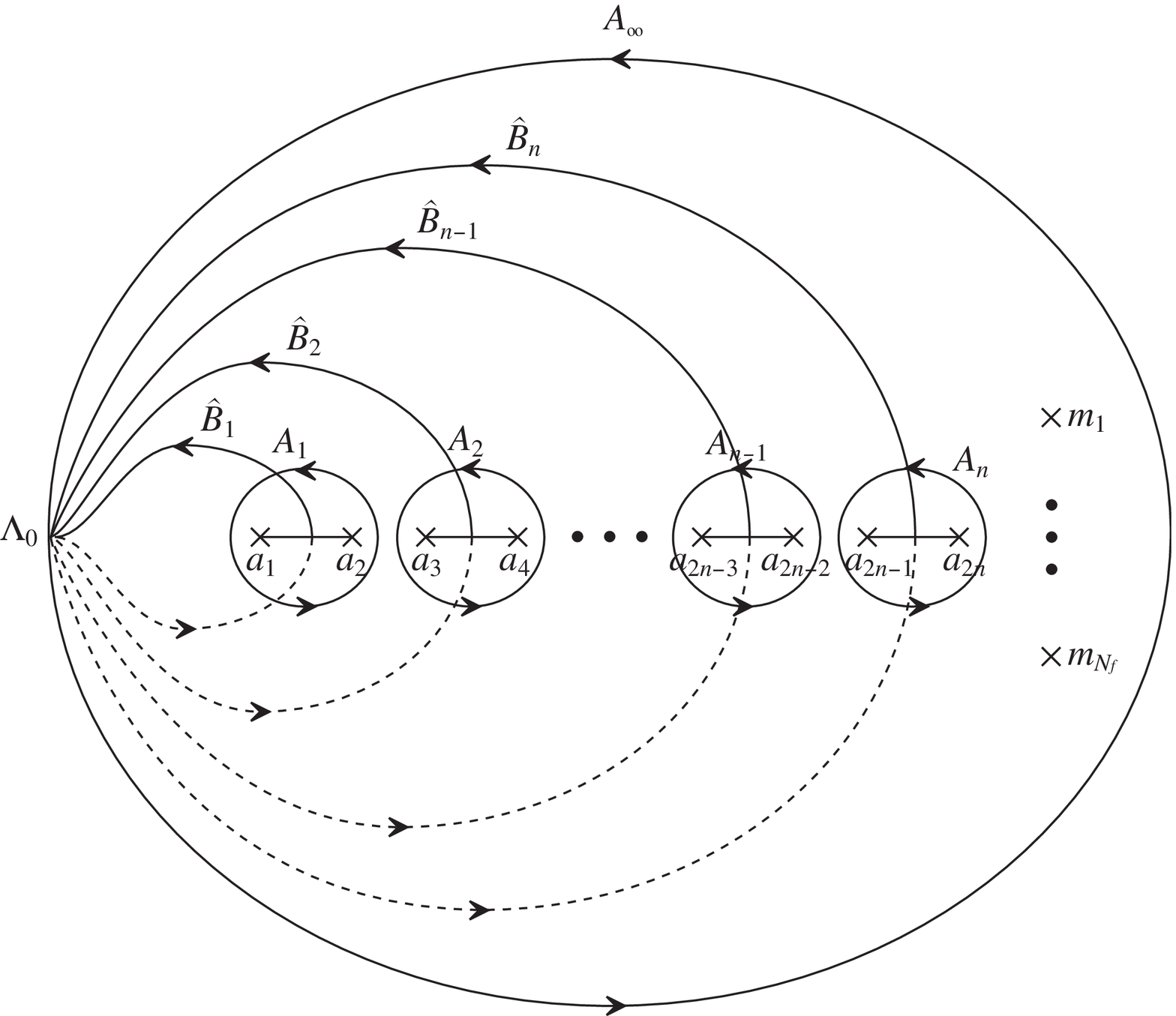}}}
\caption{The contours. Those 
on the second sheet are shown in broken lines.
$A_\infty$ is defined to go along 
$|z|=|\Lambda_0|$ counter-clockwise on the $z$-plane. }
\label{contours}      
\end{figure}

The {\it  $n$ cut solution} to the large $\hat{N}$ saddle point equation 
\begin{eqnarray}
\omega(z,g_s)=\frac{\sqrt{\prod_{i=1}^{2n}(z-a_i(g_s))}}{4\pi iS}
\sum_{j=1}^{n}\oint_{A_j} dw\frac{W'(w)+\sum_{I=1}^{N_f}\frac{g_s}{w-m_I}}
{(z-w)\sqrt{\prod_{i=1}^{2n}(w-a_i(g_s))}}
\label{omegazgs}
\end{eqnarray}
defines a meromorphic differential on ${\it \Sigma}^{g_s}_{n-1}$.
The contour surrounds all the cuts 
but not the points $w=z, m_I$ 
($I=1,\ldots,N_f$). The definitions of the contours are summarized in 
Figure \ref{contours}. 
The positions of the $2n$ branch points $a_i(g_s)$ 
$(i=1,\ldots,2n)$ are determined by the $2n$ conditions 
\begin{eqnarray}
\left(
\oint_{A_\infty} 
-\sum_{I=1}^{N_f} 
\oint_{C_I} 
\right)
dw\frac{\lambda^k(W'(w)+\sum_{I=1}^{N_f}\frac{g_s}{w-m_I})}
{\sqrt{\prod_{i=1}^{2n}(w-a_i(g_s))}}=4\pi i S\delta_{k,n}~~~(k=0,\ldots,n)
\label{1/zconditions}\end{eqnarray}
and
\begin{eqnarray}
\oint_{A_j} dw ~\omega(w,g_s) = 2\pi i\frac{S_j}S
~~~(j=1,\ldots,n).
\label{Sjconditions}
\end{eqnarray}
The first $n+1$ conditions (\ref{1/zconditions}) arise from the requirement that 
$\omega(z,g_s)$ behaves like $\frac1z$ as $z\rightarrow \infty$ on the 
first sheet, whereas the latter are the conditions for a given set of filling fractions. 
Only $n-1$ of (\ref{Sjconditions}) are independent.

 The large $\hat N$ free energy for the potential 
\begin{eqnarray}
W( \Phi , g_s)&:=&W( \Phi )+g_s\sum_{I=1}^{N_f}\log( \Phi  -m_I)
\end{eqnarray}
is given by
\begin{eqnarray}\F_0(g_s)&=&-S\int_{-\infty}^\infty d\lambda\rho(\lambda,g_s) W(\lambda, g_s)
\nonumber\\&&+
S^2\int_{-\infty}^\infty d\lambda \int_{-\infty}^\infty d\lambda' \rho(\lambda,g_s)\rho(\lambda',g_s)
\log\frac{|\lambda-\lambda'|}{\Lambda
}.
\label{F0rhorholog}\end{eqnarray}
$\Lambda$ is an arbitrary 
integration constant, which may be regarded as the physical scale 
parameter of the corresponding gauge theory up to a potential 
dependent constant.

\subsection{Annulus Diagrams}
We would now like to write (\ref{F02integral}) 
in terms of the language of Riemann surfaces. 
$\omega(\lambda,g_s)$ is expanded as
\begin{eqnarray} 
\omega(z,g_s)&=&\sum_{k=0}^\infty g_s^k \omega^{(k)}(z).\end{eqnarray}
The differential $\omega^{(1)}(z)dz$ must have zero A-periods 
because the right hand side of (\ref{Sjconditions}) does not depend on $g_s$.

Let us analyze the singularities of the differential $\omega^{(1)}(z)dz$.
Since the discontinuity of $\omega(z,g_s)$
is the eigenvalue density, it satisfies
\begin{eqnarray}
\frac1{2\pi i}\sum_{j=1}^n\oint_{A_j}dz\; \omega(z,g_s) =1.
\label{omegazgsintegral}\end{eqnarray}$\omega(z,g_s)$ also satisfies 
\begin{eqnarray}
\frac1{2\pi i}\oint_{A_\infty}dz\; \omega(z,g_s) =1
\label{omegazgsintegral2}
\end{eqnarray}
because $\omega(z,g_s)$ goes to $\frac1z$ at infinity on the first sheet.
(\ref{omegazgsintegral})(\ref{omegazgsintegral2}) imply that
$\omega(z,g_s)dz$ has a pole of first order at $z=\infty$ on the first sheet with 
residue $-1$, but otherwise it is regular everywhere else on the 
first sheet. Therefore we conclude that $\omega^{(1)}(z)dz$ is regular
everywhere on the first sheet. On the other hand,  $\omega(z,g_s)$ can be 
written in the form
\begin{eqnarray}\omega(z,g_s)&=&\frac1{2S}\left(
W'(z)+\sum_I\frac{g_s}{z-m_I}-y(z,g_s)
\right),\nonumber\\
y(z,g_s)
&=&
-\frac{\sqrt{\prod_{i=1}^{2n}(z-a_i(g_s))}}{2\pi i}
\left(\oint_{A_\infty}-\sum_{I=1}^{N_f}\oint_{C_I}\right)
dw\frac{W'(w)
+\sum_{I=1}^{N_f}\frac{g_s}{\lambda-m_I}}
{(z-w)\sqrt{\prod_{i=1}^{2n}(w-a_i(g_s))}}.
\nonumber \\
&&
\label{yzgs}\end{eqnarray}
We similarly expand
\begin{equation} 
y(z,g_s)=\sum_{k=0}^\infty g_s^k y^{(k)}(z).\end{equation}
Note that the difference of the contours of $y(z,g_s)$ and 
$\omega(z,g_s)$; thanks to this change, the contour integral of 
$y(z,g_s)$ gives some rational function of $z$.\footnote{If $g_s=0$, it is a polynomial of $z$. If, in particular, $n$ saturates 
the degree of $W'(z)$ (the maximally broken case), it is a constant up to which $y^{(0)}(z)$ coincides with $y(z)$ (\ref{y}).\label{footnote1}}
This implies that 
$y(z,g_s)$ changes its sign under the interchange of  
the first and second sheets.

Since the second term of $\omega(z,g_s)$ 
has a first order singularity at each $z=m_I$ 
$(I=1,\ldots,N_f)$ with residue $\frac{g_s}{2S}$
and at $z=\infty$ with residue $-\frac{g_s N_f}{2S}$ on both sheets, 
the $y(z,g_s)$ term must cancel them on the first sheet, and hence 
in turn doubles the residues on the second sheet. To summarize, 
denoting coordinates on the second sheet with tildes, 
$\omega^{(1)}(z)dz$ has 
first order singularities at 
\begin{itemize}
\item[$\bullet$] $z=\tilde{m}_I$ with residue $\frac{1}{S}$ for all $I=1,\ldots,N_f$, and
\item[$\bullet$] $z=\tilde\infty$ with residue $-\frac{N_f}{S}$,
\end{itemize}
otherwise holomorphic everywhere else on ${\it \Sigma}^{g_s}_{n-1}$. 
Thus we have shown that
$\omega^{(1)}(z) dz$ is an abelian differential of the 
third kind with zero A-periods and is given by
\begin{eqnarray}\omega^{(1)}(z) dz =
\frac1S \sum_{I=1}^{N_f}\omega_{\tilde m_I -\tilde\infty},\end{eqnarray}
where, following the notation used in \cite{Fay},
$\omega_{a-b}(z)$ denotes a zero A-period abelian differential 
which has simple poles at $z=a,b$ with residue $\pm1$, respectively.

One of the nice things for this kind of differentials is that 
their integrals are compactly described in terms of the 
prime form $E(z,w)$ on ${\it \Sigma}^{g_s}_{n-1}$.      
The prime form $E(z,w)$ is known to be the unique bi-holomorphic half 
differential on a compact Riemann surface such that  $E(z,w)=0$ if 
and only if 
$z=w$. One of the basic properties of 
the prime form is \cite{Fay}
\begin{eqnarray}\omega_{a-b}(z)&=&d_z\log\frac{E(z,a)}{E(z,b)}.\end{eqnarray}
Using this formula, we finally obtain the third type of contributions to 
the gravitational correction\footnote{Of course, the computation of the annulus amplitude 
$\F_0^{(2)}$ itself can also be (and has been) done by other means
(See e.g. \cite{AJM,IK}.). What is 
new here is that we have expressed it in terms of a special form (that is, the prime 
form) on the Riemann surface, and we cannot compare it with 
the CFT result until we do so. We thank H. Kawai for discussions on this
point.}  
\begin{eqnarray}\F_0^{(2)}&=&-\frac{1}{4\pi i}
\sum_{j=1}^n
\oint_{A_j}  
\sum_{J=1}^{N_f}
\omega_{\tilde m_J-\tilde\infty}(z) 
\sum_{I=1}^{N_f}
\log(z-m_I)
\nonumber\\
&=&
\frac{1}{2}\sum_{I,J=1}^{N_f}
\int_{m_I}^\infty  \omega_{\tilde m_J-\tilde\infty}(z) 
\nonumber\\
&=&
\frac12\sum_{I,J=1}^{N_f}
\log\frac{E(m_I,\tilde\infty)E(\infty,\tilde m_J)}{E(m_I,\tilde m_J)E(\infty,\tilde\infty)}.
\label{F0(2)primeform}
\end{eqnarray}
The full planar gravitational correction $\F^{(\mbox{\scriptsize pl})}$ 
is thus
\begin{eqnarray}
\F^{(\mbox{\scriptsize pl})}&=&
\frac12\sum_{i,j=1}^n N_i N_j \frac{\partial^2 
\F^{(0)}_0
}{\partial S_i \partial S_j}
+\sum_{i=1}^n N_i \frac{\partial 
\F^{(1)}_0
}{\partial S_i }+\F^{(2)}_0 \label{Fplgr}
\end{eqnarray}
with 
\begin{eqnarray}
\frac12\sum_{i,j=1}^n N_i N_j\frac{\partial^2\F_0^{(0)}}{\partial S_i\partial S_j}&=&
\lim_{|\Lambda_0|\rightarrow\infty}\left[ 
\sum_{i,j=1}^{n}N_i N_j\left(
-\pi i 
\int_{\hat B_i} \hat\omega_j(z)
+\log \frac{-\Lambda_0}{\Lambda}
\right)
\right],
\label{dF2dSdS}\\
\sum_{i=1}^nN_i
\frac{\partial \F_0^{(1)}}{\partial S_i}
&=&-
\sum_{i,k=1}^nN_i\oint_{A_k}
\hat\omega_i(z) \sum_{I=1}^{N_f} \log(z-m_I)
\nonumber\\
&=&\lim_{|\Lambda_0|\rightarrow\infty}\left[\rule{0mm}{7mm}\right.
\sum_{i=1}^nN_i
\sum_{I=1}^{N_f} 
\left(2\pi i
\int_{m_I}^{\Lambda_0} 
\hat\omega_i(z)
-\log(-\Lambda_0)
\right)
\left.\rule{0mm}{7mm}\right]
\label{NdFdS}
\end{eqnarray}
and $\F_0^{(2)}$ given in (\ref{F0(2)primeform}). 
In deriving the above, 
we have used the special geometry relation \cite{DV,CSW}
\begin{eqnarray}\frac{\partial \F_0^{(0)}}{\partial S_i}&=&
\lim_{|\Lambda_0|\rightarrow\infty}\left(
\frac12\int_{\hat B_i}dz\; y^{(0)}(z)
-W(\Lambda_0)+ 2S \log \frac{-\Lambda_0}{\Lambda}
\right)
\end{eqnarray}
as well as the variation formula
\begin{eqnarray}
\frac{\partial y^{(0)}(z)}{\partial S_i}dz&=&-4\pi i \hat \omega_i(z)
(i=1,\ldots,n),  
\label{dy0dS}
\end{eqnarray}
where $\hat\omega_i$ $(i=1,\ldots,n)$ are related with the 
holomorphic differentials $\omega_i$ $(i=1,\ldots,n-1)$ on 
${\it \Sigma}^{g_s=0}_{n-1}$ through the relations 
\begin{eqnarray}\hat\omega_i&=&\omega_i-\frac1{2\pi i}\omega_{\infty-\tilde\infty}~~~
(i=1,\ldots,n-1),\nonumber\\
\hat\omega_n&=&-\frac1{2\pi i}\omega_{\infty-\tilde\infty}.
\end{eqnarray}
They are normalized so that 
\begin{eqnarray}
\oint_{A_i} \omega_j(z) &=&\delta_{ij}~~~(i,j=1,\ldots,n-1),\nonumber\\
\oint_{A_i} \hat\omega_j(z) &=&\delta_{ij}~~~(i,j=1,\ldots,n).
\end{eqnarray}
See Appendix B for details.

\section{Examples} 
\subsection{One-cut Solutions for Quadratic Potential with $N_f=1$}
The first example is the `conifold' case
\begin{eqnarray}
W( \Phi )&=&-t  \Phi ^2
\end{eqnarray}
with a single matter field.
The resolvent $\omega(z,g_s)$ is 
\begin{eqnarray}\omega(z,g_s)&=&\frac1{2S}
\left[\rule{0mm}{7mm}
-2tz+\frac{g_s}{z-m}\right.\nonumber\\
&&\left.+\sqrt{(z-c(g_s))^2-\mu(g_s)^2}
\left(
2t-\frac{\frac{g_s}{z-m}}{\sqrt{(m-c(g_s))^2-\mu(g_s)^2}}
\right)
\right],\nonumber\\\end{eqnarray}where $c(g_s)$ and $\mu(g_s)$ are related with the positions of the 
end points by
\begin{eqnarray}
a_1(g_s)&=&c(g_s)+\mu(g_s),\nonumber\\
a_2(g_s)&=&c(g_s)-\mu(g_s)
\end{eqnarray}
and satisfy  (\ref{1/zconditions}) and (\ref{Sjconditions})    
\begin{eqnarray}
0&=&2tc(g_s)+\frac{g_s}{\sqrt{(m-c(g_s))^2-\mu(g_s)^2}},\\
2S&=&-t \mu(g_s)^2+g_s\left(
1-\frac{m-c(g_s)}{\sqrt{(m-c(g_s))^2-\mu(g_s)^2}}
\right).
\end{eqnarray} 
They are solved  as a power series
\begin{eqnarray}c(g_s)&=&
 -\frac1{2t\sqrt{m^2-\mu^2}}
 g_s 
 + 
 \frac{-\sqrt{m^2-\mu^2}+2m}{2t^2(m^2-\mu^2)^2}
 g_s^2 
 +{\cal O}(g_s^3),
\label{cquadratic}
\\
\mu(g_s)&=&
\mu
+ 
\frac{\sqrt{m^2-\mu^2}-m}{2t\mu\sqrt{m^2-\mu^2}}
g_s  + 
\frac{-(m^2-2\mu^2-m\sqrt{m^2-\mu^2})^2+5 \mu^4}
{4t^2\mu^3(m^2-\mu^2)^2}
g_s^2  
+{\cal O}(g_s^3)\nonumber\\
\label{muquadratic}
\end{eqnarray}
with $\mu:=\sqrt{-\frac{2S}t}$.
The integral for the genus zero free energy $\F_0$ (\ref{F0rhorholog}) can be 
completely performed in this case and is given in Appendix C. 
 $\F_0^{(1)}$ (the disk amplitude) and $\F_0^{(2)}$ (the annulus amplitude) 
are obtained as coefficients of the Taylor series of $\F_0(g_s)$. 
After some calculations 
using (\ref{cquadratic}) and (\ref{muquadratic})
we find 
\begin{eqnarray}\F_0^{(1)}&=&-S\left(
\log\frac{m+\sqrt{m^2-\mu^2}}2+
\frac m{m+\sqrt{m^2-\mu^2}}-\frac12
\right),
\label{F0(1)onecut}
\\
\F_0^{(2)}&=&\frac12 \log\frac{m+\sqrt{m^2-\mu^2}}{2\sqrt{m^2-\mu^2}}
~~~~~~(\mu={\textstyle \sqrt{-\frac{2S}t}}).
\label{F0(2)conifold}
\end{eqnarray}
$\F_0^{(1)}$ (\ref{F0(1)onecut}) agrees with the calculations 
in \cite{matter}.

Let us show that (\ref{F0(2)conifold}) can be written in terms of prime forms. 
Let $w,w'\in\mathbb{C}\cup\{\infty\}={\bf P}^1$ be coordinates of some two points 
in the natural coordinate system, then
the prime form is simply given by
\begin{eqnarray}E(w,w')
&=&\frac{w-w'}
{\sqrt{dw}\sqrt{dw'}}.
\end{eqnarray}
${\bf P}^1$ is also realized as a two-sheeted Riemann surface with a single 
branch cut. We take two end points at 
$z=\pm\mu$ ($\mu\in\mathbb{R}$, $\mu>0$), 
where $z$ 
is the coordinate of the two-sheeted Riemann surface. The relation between
these two coordinate systems is 
\begin{eqnarray}z=w+\frac{\mu^2}{4w},&&w=\frac{z+\sqrt{z^2-\mu^2}}2.\end{eqnarray} 
This map is two to one for generic $z$, except at the two end points of the 
cut where it is one to one.

The cut ${[}-\mu,\mu{]}$ on the two-sheeted $z$ plane
gets mapped to a circle $|w|=\frac\mu2$ on the $w$ plane.
Defining the first and second sheets as before, 
the region outside this circle corresponds to the first sheet, while inside 
the second sheet.   
It is easy to see that
\begin{eqnarray}
E(z,\tilde z)
&=&\frac{w-\frac{\mu^2}{4w}}
{\sqrt{dw(z)}
\sqrt{dw(\tilde z)}}
\nonumber\\
&=&\frac{\sqrt{z^2-\mu^2}
}
{\sqrt{dw(z)}
\sqrt{dw(\tilde z)}},
\nonumber\\
E(z,\tilde \infty)
&=&\frac{w-0}{\sqrt{dw(z)
dw(\tilde\infty)}}
\nonumber\\
&=&\frac{\frac{z+\sqrt{z^2-\mu^2}}2} 
{\sqrt{dw(z) 
dw(\tilde\infty)}},\nonumber\\
E(\infty,\tilde z)
&=&\frac{\infty-\frac{\mu^2}{4w}}
{\sqrt{dw(\infty) 
dw(\tilde z)}},\nonumber\\
E(\infty,\tilde \infty)
&=&\frac{\infty-0}{\sqrt{dw(\infty) 
dw(\tilde \infty)}},
\end{eqnarray}
where, as before, we have added tildes to the $z$ coordinates
for the points on the second sheet.
Thus one may write (\ref{F0(2)conifold}) as
\begin{eqnarray}
\F_0^{(2)}&=&\frac12\log\frac{E(m,\tilde \infty)E(\infty,\tilde m)}
{E(m,\tilde m)E(\infty,\tilde \infty)}.
\end{eqnarray}
This agrees with our general formula (\ref{F0(2)primeform}).

\subsection{Two-cut Solutions for a Quartic Potential with $N_f=2$}
The next example is an even quartic superpotential with two flavors 
with masses 
\begin{eqnarray}
m_1=-m_2:= m, 
\end{eqnarray}
in which 
we find a symmetric $2$-cut solution.
The potential is
\begin{eqnarray}
W(\Phi)=-t \Phi^2-u \Phi^4.
\label{quartic pot}
\end{eqnarray}
Without matter, a large $N$ symmetric solution was obtained long time 
ago by \cite{Shimamune} in terms of elementary functions. Despite 
the symmetric potential, asymmetric filling is also possible and the solution 
is written using elliptic functions in general \cite{Ferrari,Shih}.

After integrating out the matter fields, we have 
\begin{eqnarray}
W(\lambda, g_s)=-t \lambda^2-u \lambda^4 +g_s \log (\lambda^2-m^2).
\nonumber
\end{eqnarray}
With a symmetric ansatz, we find the resolvent 
\begin{eqnarray}
&&\hspace*{-0.7cm}
2S \omega(z,g_s)=-2tz-4uz^3+\frac{g_s}{z-m}+\frac{g_s}{z+m}
\nonumber \\
&&\hspace*{1.7cm}
+z\sqrt{(z^2-a^2)(z^2-b^2)}\left[
4 u-\frac{2 g_s}{(z^2-m^2)\sqrt{(m^2-a^2)(m^2-b^2)}}
\right],
\label{2-cut resol2}
\end{eqnarray}
where two cuts are created at $[-b,-a]$ and $[a,b]$.
The density function of eigenvalues yields
\begin{eqnarray}
\rho(\lambda,g_s)=\frac{1}{2 \pi S}\lambda
\sqrt{(\lambda^2-a^2)(b^2-\lambda^2)}\left[
-4 u+\frac{2 g_s}{(\lambda^2-m^2)\sqrt{(m^2-a^2)(m^2-b^2)}}
\right].
\label{2-cut density}
\end{eqnarray}

Being symmetric, the condition (\ref{Sjconditions}) is automatically 
satisfied, while the asymptotic conditions (\ref{1/zconditions}) yield  
\begin{eqnarray} 
&&0=2uc+t+\frac{g_s}{\sqrt{(c-M)^2-\mu^2}},
 \\
&& S=
-u\mu^2+g_s\left(1-
\frac{M-c}{\sqrt{(M-c)^2-\mu^2}}
\right),
\end{eqnarray}
where parameters $c$, $\mu$ and $M$ are defined by
\begin{eqnarray}
&&c=\frac{a+b}{2}, \quad
\mu=\frac{b-a}{2}, \quad
M=m^2.
\nonumber
\end{eqnarray}
From these conditions 
we find 
an iterative solution
\begin{eqnarray}
&&c=-\frac{t}{2u}-\frac{g_s}{\sqrt{(t+2uM)^2+4uS}}
+\frac{2 g_s^2 u(2t+4uM-\sqrt{(t+2uM)^2+4uS})}
{[(t+2uM)^2+4uS]^2}+{\cal O}(g_s^3),
\nonumber \\
&&\mu=
\sqrt{-\frac{S}{u}}+\frac{g_s}{2}\sqrt{-\frac{1}{uS}}
\left(
1-\frac{t+2uM}{(t+2uM)^2+4uS}
\right)
\nonumber \\
&&\hspace*{1cm}
+\frac{g_s^2}{2}\sqrt{-\frac{u}{S}}
\Biggl[
-\frac{2}{(t+2uM)^2+4uS}
+\frac{1}{4uS}\left(
1-\frac{t+2uM}{\sqrt{(t+2uM)^2+4uS}}
\right)^2
\nonumber \\
&&\hspace*{1cm}
+\frac{2(t+2uM)(2t+4uM-\sqrt{(t+2uM)^2+4uS})}
{((t+2uM)^2+4uS)^2}
\Biggr]+{\cal O}(g_s^3).
\label{2-cut center}
\end{eqnarray}

The multi-cut large $\hat N$ free energy is given in Appendix A, 
and is in the present case 
\begin{eqnarray}
\F_0&=&-S\int d\lambda \rho(\lambda,g_s)\left(
\frac{1}{2}W(\lambda, g_s)-\frac{S}{2}\log \frac{|\lambda^2-\lambda_0^2|}
{\Lambda^2}
\right)
-\frac{S}{2}W(\lambda_0,g_s),
\label{2-cut free}
\end{eqnarray}
where $\int_{-b}^{-a}d\lambda \rho(\lambda,g_s)
=\int_{a}^b d\lambda \rho(\lambda,g_s)=1/2$.
Plugging (\ref{2-cut density}) (\ref{2-cut center}) into (\ref{2-cut free}) 
and picking up the ${\cal O}(g_s^2)$ terms, we obtain the free energy of annulus 
diagrams as before.
The result is
\begin{eqnarray}
\F_0^{(2)}=\frac{1}{2}\log \left(
\frac{t+2uM+\sqrt{(t+2uM)^2+4uS}}{2 \sqrt{(t+2uM)^2+4uS}}
\right).
\end{eqnarray}

Let us compare this with (\ref{F0(2)primeform}).
The prime form is defined on the curve
\begin{eqnarray}
&&y^2=(z^2-a^2)(z^2-b^2),
\nonumber \\
&&a^2=-\frac{t}{2u}-\sqrt{-\frac{S}{u}},
\quad
b^2=-\frac{t}{2u}+\sqrt{-\frac{S}{u}}.
\end{eqnarray}
In the present case 
(\ref{F0(2)primeform}) reads
\begin{eqnarray}
\frac12\log\frac{E(m,\tilde{m})E(m,-\tilde{m})E(-m,\tilde{m})E(-m,-\tilde{m})}
{E(m,\tilde{\infty})^2E(-m,\tilde{\infty})^2}
&=&\frac12\left(\int_m^\infty + \int_{-m}^\infty\right)
(\omega_{\tilde m-\tilde \infty}+
\omega_{(-\tilde m)-\tilde \infty}).
\nonumber\\
\label{F02Nf=2}
\end{eqnarray}
The relevant meromorphic 1-form 
with zero A-periods is 
\begin{eqnarray}
\omega_{\tilde m-\tilde \infty}+
\omega_{(-\tilde m)-\tilde \infty}
=-\frac{1}{2}\left(
\frac{y(m)}{y(z)}\frac{1}{z-m}+\frac{y(-m)}{y(z)}\frac{1}{z+m}
-\frac{1}{z-m}-\frac{1}{z+m}
\right)dz-\frac{zdz}{y(z)}.
\nonumber\\
\end{eqnarray}
By changing the integration variable $z$ to $x=z^2$,
the above integration reduces to a simple form\footnote{
Identifying $z\sim-z$, we obtain the same curve and singular points 
as those in the conifold case. 
}
\begin{eqnarray}  
(\ref{F02Nf=2})&=&\frac{1}{2}\int_{M}^{\infty}dx 
\left\{
\frac{1}{x-M}-\frac{y(M)}{y(x)}\frac{1}{x-M}
-\frac{1}{y(x)}
\right\}
\nonumber \\
&=&
\frac{1}{2}\log \left(
\frac{t+2uM+\sqrt{(t+2uM)^2+4uS}}{2 \sqrt{(t+2uM)^2+4uS}}
\right).
\end{eqnarray}
This result exactly coincides with the matrix model calculation.

\subsection{Two-cut Solutions for a Cubic Potential with $N_f=1$}
\subsubsection{Perturbative Computations in Gauged Matrix Models} 
The final example is the perturbative computations 
of the two-cut free energy for a cubic potential
\cite{DGKV}
\begin{eqnarray}
W=g{\rm Tr}\left(
\frac{1}{3}\Phi^3+\frac{\Delta}{2}\Phi^2
\right)+\tilde{Q}(\Phi-m)Q.
\end{eqnarray}
In this  two-cut case 
we need to consider the fluctuation around the vacuum
\begin{eqnarray}
\Phi=\left(
\begin{array}{cc}
a_1{\bf 1}_{\hat{N}_1} & 0 \\
0 & a_2{\bf 1}_{\hat{N}_2}
\end{array}
\right)+
\left(
\begin{array}{cc}
\Phi_{11} & \Phi_{12} \\
\Phi_{21} & \Phi_{22}
\end{array}
\right).
\end{eqnarray}
where $a_1$ and $a_2$ are classical roots of $W^{\prime}(z)=0$.
Around this vacuum the original gauge symmetry 
of matrix model 
\begin{eqnarray}
\Phi\to U\cdot \Phi \cdot U^{-1},  \quad U\in U(\hat{N})
\end{eqnarray}
reduces to $U(\hat{N})\to U(\hat{N}_1)\times U(\hat{N}_2)$,
and the matrix model action is given by
\begin{eqnarray}
W_{\rm tree}(\Phi)=g{\rm Tr}\left(
\frac{1}{3}\Phi_{11}^3+\frac{\Delta}{2}\Phi_{11}^2
\right)
+g{\rm Tr}\left(
\frac{1}{3}\Phi_{22}^3-\frac{\Delta}{2}\Phi_{22}^2
\right).
\end{eqnarray}
Since the off-diagonal block of the matrix does not appear
in the action, a convenient gauge choice is
\begin{eqnarray}
\Phi_{ij}=0, \quad i\ne j.
\end{eqnarray}
In this gauge the coupling to vectors $Q, \tilde{Q}$ is
\begin{eqnarray}
W_{\rm matter}=\tilde{Q}_1(\Phi_{11}-m)Q_{1}
+\tilde{Q}_2(\Phi_{22}-m-\Delta)Q_{2}.
\end{eqnarray}
The gauge fixing requires the introduction of the ghost matrices 
$B$ and $C$ with action 
\begin{eqnarray}
W_{\rm ghost}(B,C)&=&
\Delta {\rm Tr}(B_{21}C_{12})-\Delta {\rm Tr}(B_{12}C_{21})
\nonumber \\
&&+{\rm Tr}(B_{21}\Phi_{11}C_{12}+C_{21}\Phi_{11}B_{12})
\nonumber \\
&&+{\rm Tr}(B_{12}\Phi_{22}C_{21}+C_{12}\Phi_{22}B_{21}).
\end{eqnarray}

\begin{figure}
\centering
\resizebox{0.75\textwidth}{!}{%
  \includegraphics{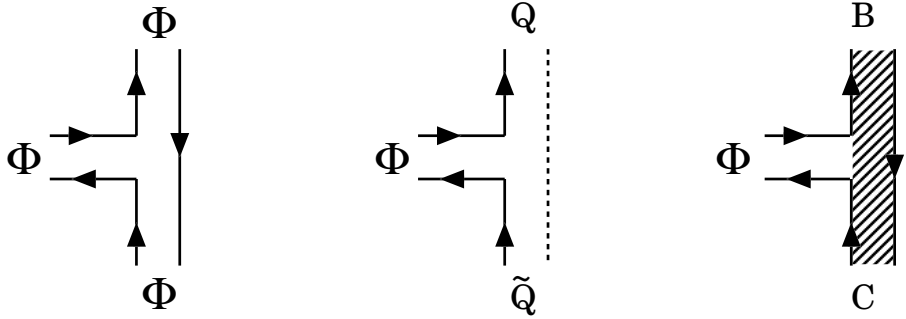}}
\caption{Vertices in the gauged matrix model.}
\label{gh}      
\end{figure}

From the matrix model action 
\begin{eqnarray}
W(\Phi,B,C)&=&W_{\rm tree}(\Phi)+W_{\rm ghost}(B,C),
\end{eqnarray}
we can read off the propagators
\begin{eqnarray}
&&\langle \Phi_{11}\Phi_{11}\rangle=\frac{1}{g\Delta}, \quad
\langle \Phi_{22}\Phi_{22}\rangle=-\frac{1}{g\Delta},
\nonumber \\
&&\langle B_{12}C_{21}\rangle=-\langle B_{21}C_{12}\rangle
=\frac{1}{g\Delta},
\nonumber \\
&&\langle Q_{1}Q_{1}\rangle=-\frac{1}{m}, \quad
\langle Q_{2}Q_{2}\rangle=-\frac{1}{m+\Delta}.
\end{eqnarray}
For $\Phi^3$ and $B\Phi C$ vertices we assign weight $g$,
and for $\tilde{Q}\Phi Q$ vertex  we assign weight $1$ 
(Figure \ref{gh}).
For a ghost loop, we add an extra factor of $-1$.

\begin{figure}
\centering
\resizebox{0.35\textwidth}{!}{%
  \includegraphics{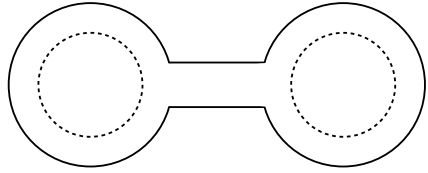}}
\caption{The dump-bell diagram.}
\label{dumpbell2}      
\end{figure}

\begin{figure}
\centering
\resizebox{0.75\textwidth}{!}{%
  \includegraphics{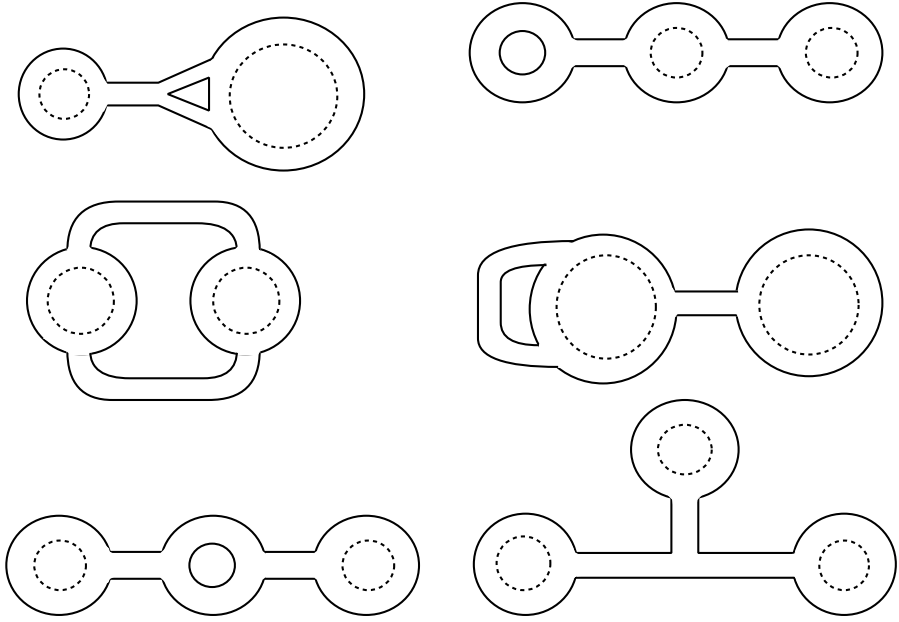}
} 
\caption{Three-loop diagrams contributing to $S_1^2$ or $S_2^2$.}
\label{3-loop1}   
\end{figure}

\begin{figure}
\centering
\resizebox{0.75\textwidth}{!}{%
  \includegraphics{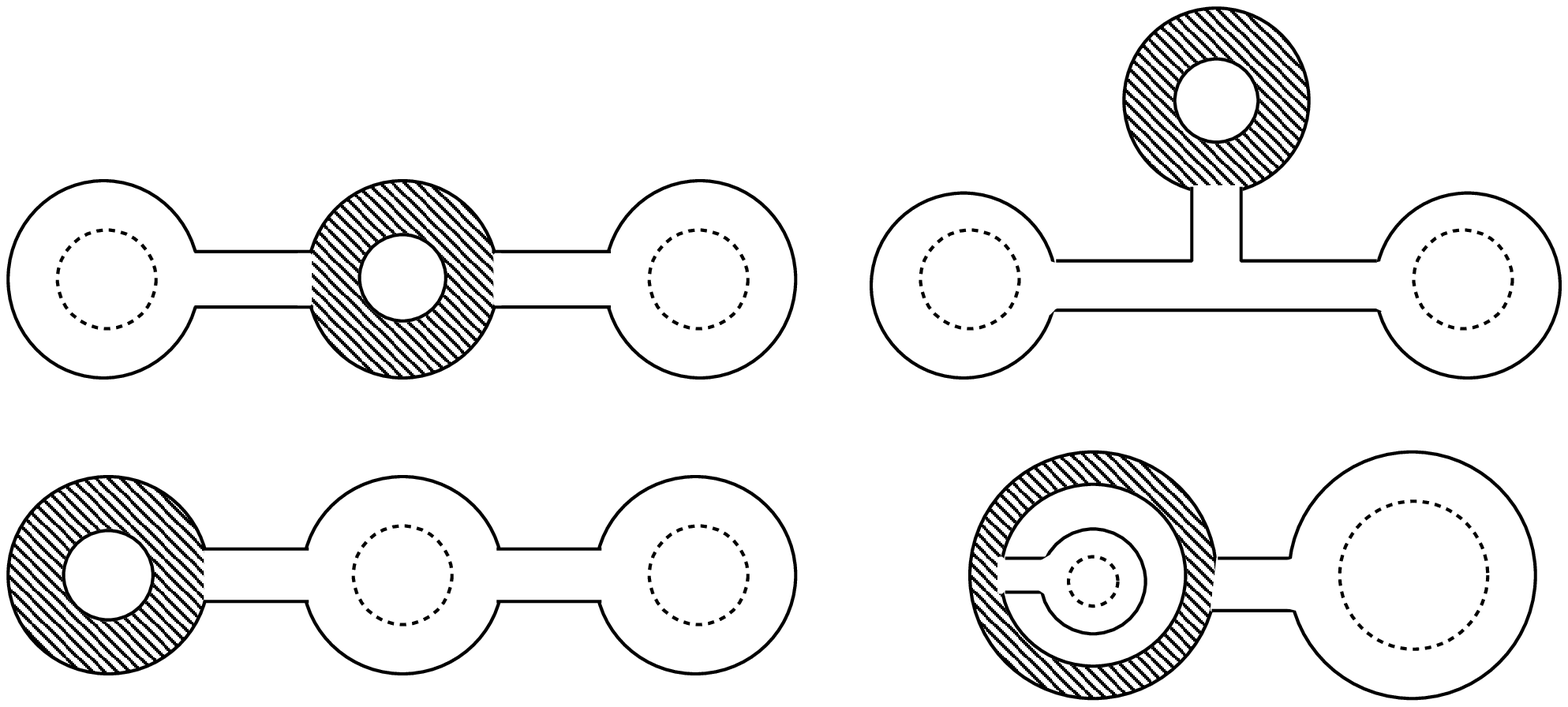}
}
\caption{Three-loop diagrams contributing to $S_1S_2$.}
\label{3-loop2}   
\end{figure}

Having found the Feynman rules, 
we can calculate the annulus contributions perturbatively.
The annulus diagrams are drawn in 
Figures \ref{dumpbell2}-\ref{3-loop2} up to three loops,.
The contributions to annulus free energy 
from the dump-bell diagram Figure \ref{dumpbell2} is
\begin{eqnarray}
&&\frac{S_1}{2g\Delta m^2}
-\frac{S_2}{2g\Delta (m+\Delta)^2}.
\end{eqnarray}

From the three-loop diagrams, we obtain the terms 
involving the $S_1^2$, $S_2^2$ and $S_1S_2$.
The diagrams in Figure \ref{3-loop1}  
lead to the $S_1^2$ and $S_2^2$ terms.
We obtain the $S_1^2$ terms 
\begin{eqnarray}
\Bigl[
-\frac{1}{g^2\Delta^3m^3}-\frac{1}{g^2\Delta^3m^3}
+\frac{1}{4g^2\Delta^2m^4}+\frac{1}{g^2\Delta^2m^4}
+\frac{1}{g^2\Delta^4m^2}+\frac{1}{g^2\Delta^4m^2}
\Bigr]S_1^2.
\label{Graph1}
\end{eqnarray}
The $S_2^2$ terms can be obtained by replacements
$S_1\to S_2$, 
$\Delta \to -\Delta$ and 
$m\to m+\Delta$ 
\begin{eqnarray}
&&\Bigl[
\frac{1}{g^2\Delta^3(m+\Delta)^3}+\frac{1}{g^2\Delta^3(m+\Delta)^3}
+\frac{1}{4g^2\Delta^2(m+\Delta)^4}+\frac{1}{g^2\Delta^2(m+\Delta)^4}
\nonumber \\
&&+\frac{1}{g^2\Delta^4(m+\Delta)^2}+\frac{1}{g^2\Delta^4(m+\Delta)^2}
\Bigr]S_2^2.
\label{Graph2}
\end{eqnarray}
Finally, the diagrams in Figure \ref{3-loop2} lead to the $S_1S_2$ terms.
\begin{eqnarray}
&&-\Biggl[
\frac{1}{g^2\Delta^4}
\left(
\frac{1}{m^2}+\frac{1}{(m+\Delta)^2}
\right)
+\frac{2}{g^2\Delta^4}
\left(
\frac{1}{m^2}+\frac{1}{(m+\Delta)^2}
\right)
\nonumber \\
&&\hspace*{0.7cm}
+\frac{2}{g^2\Delta^3}
\left(
-\frac{1}{m^3}+\frac{1}{(m+\Delta)^3}
\right)
+\frac{2}{g^2\Delta^4m(m+\Delta)}
\Biggr]S_1S_2.
\label{Graph3}
\end{eqnarray}

\subsubsection{Comparison with the Annulus Formula}
To compare these diagrammatic computation with the 
annulus formula (\ref{F0(2)primeform}), let us expand the 
latter in powers of $S_1$ and $S_2$. 
For the two-cut solution the prime form is defined on 
the curve
\begin{eqnarray}
y^2&=&\left(\frac{1}{g}W^{\prime}(z)\right)^2+f_{1}(z)
=(z-z_1)(z-z_2)(z-z_3)(z-z_4),
\nonumber \\
\label{2-cut cub curve}
W(z)&=&g\left(
\frac{1}{3}z^3+\frac{\Delta}{2}z^2
\right).
\end{eqnarray}
There are two 1-cycles :  
$A_1$ around $[z_1,z_2]$ and 
$A_2$ around $[z_3,z_4]$.
We assume that the two cuts to be small so that we shall find 
a solution in a power series of the periods.

Using 
the addition theorem for the prime form,  we write  
the annulus formula (\ref{F0(2)primeform}) 
on the curve (\ref{2-cut cub curve}) as
\begin{eqnarray}
\frac{1}{2}\log\frac{E(m,\tilde{m})E(\tilde\infty,\infty)}
{E(m,\tilde{\infty})E(\tilde{m},\infty)}
=\frac{1}{2}\int_{m}^{\infty}\omega_{\tilde{m}-\tilde{\infty}}.
\label{cubic prime frac}
\end{eqnarray}
The meromorphic 1-form $\omega_{\tilde{m}-\tilde{\infty}}$ 
on this curve is given by
\begin{eqnarray}
\omega_{\tilde{m}-\tilde{\infty}}
=
\frac{1}{2}\left[
\frac{dz}{z-m}
\left(
1-\frac{y(m)}{y(z)}
\right)
+\frac{(\alpha_1z+\alpha_2)dz}{y(z)}\right].
\end{eqnarray}
The coefficients $\alpha_1$, $\alpha_2$ for the holomorphic forms 
are determined by the zero A-period conditions
\begin{eqnarray}
&&\sum_{i=1}^2\oint_{A_i}\omega_{\tilde{m}-\tilde{\infty}}=0,
\nonumber \\
&&\oint_{A_2}\omega_{\tilde{m}-\tilde{\infty}}=0.
\nonumber
\end{eqnarray}
The integration in the first condition can be performed 
by deforming the contour to $A_\infty\cup (-{\cal C}_{m})$.
As a result, we obtain $\alpha_1=1$.
The integration in the second condition is rather difficult.
To perform this integration, we introduce the parameterization 
\cite{CIV}
\begin{eqnarray}
&&
z_1=\frac{Q-I}{2}-\Delta_{21}, \quad 
z_2=\frac{Q-I}{2}+\Delta_{21},
\nonumber \\
&&
z_3=\frac{Q+I}{2}-\Delta_{43}, \quad 
z_4=\frac{Q+I}{2}+\Delta_{43},
\nonumber \\
&&
Q:=\frac{z_1+z_2+z_3+z_4}{2},
\quad
I:=\frac{z_3+z_4}{2}-\frac{z_1+z_2}{2}.
\nonumber
\end{eqnarray}
By expanding the condition 
$\int_{z_3}^{z_4}\omega_{\tilde{m}-\tilde{\infty}}=0$
in terms of $\Delta_{21}$, we determine the coefficient 
$\alpha_2$ perturbatively up to ${\cal O}(\Delta_{21}^4)$ as
\begin{eqnarray}
\alpha_2&=&\sqrt{(m^{\prime}-z_3)(m^{\prime}-z_4)}+
\frac{\Delta_{21}^2(z_3+z_4)\biggl(-m^{\prime}
+\sqrt{(m^{\prime}-z_3)(m^{\prime}-z_4)}\biggr)}
{4z_3z_4m^{\prime}}
\nonumber \\
&&+\frac{\Delta_{21}^4}{128 z_3^3z_4^3m^{\prime\;3}}
\biggl[
-m^{\prime\;2}\biggl(m^{\prime}+\sqrt{(m^{\prime}-z_3)(m^{\prime}-z_4)}
\biggr)
(z_3+z_4)(9z_3^2-10z_3z_4+9z_4^2)
\nonumber \\
&&
+\sqrt{(m^{\prime}-z_3)(m^{\prime}-z_4)}\biggl(
8z_3^2z_4^2(z_3+z_4)
+2z_3z_4m^{\prime}(3z_3^2+2z_3z_4+3z_4^2)
\biggr)
\biggr]+{\cal O}(\Delta_{21}^6),
\nonumber 
\end{eqnarray}
where $m^{\prime}:= m+\frac{I-Q}{2}$.

The periods of the curve ($\ref{2-cut cub curve}$)
\begin{eqnarray}
S_1=g \int_{z_3}^{z_4}\frac{dz}{2\pi}\;y(z),
\quad
S_2=g \int_{z_1}^{z_2}\frac{dz}{2\pi}\;y(z)
\end{eqnarray}
are evaluated in terms of the variables 
$(Q,I,\Delta_{21},\Delta_{43})$.
We can iteratively solve the inverse relations 
$\Delta_{21}(S_1,S_2,g,\Delta)$, $\Delta_{43}(S_1,S_2,g,\Delta)$
as \cite{KMT}
\begin{eqnarray}
&&\Delta_{43}(S_1,S_2,g,\Delta)^2=\frac{4}{g\Delta}S_1
+\frac{8}{g^2\Delta^4}(2S_1^2-3S_1S_2)
+\frac{8}{g^3\Delta^7}S_1(5S_1-13S_2)(4S_1-3S_2)
\nonumber \\
&&\hspace*{3.7cm}
+{\cal O}(S^4),
\nonumber \\
&&\Delta_{21}(S_1,S_2,g,\Delta)^2=\Delta_{43}(S_1,S_2,g,-\Delta). 
\label{cub inv}
\end{eqnarray}

The integration of meromorphic 1-form (\ref{cubic prime frac})
can be performed by expanding in terms of $\Delta_{21}$ and 
$\Delta_{43}$. By plugging the inverse relations (\ref{cub inv}),
we obtain the perturbative expansion 
\begin{eqnarray}
\F_0^{(2)}&=&
\frac{S_1}{2g\Delta m^2}-\frac{S_2}{2g\Delta (m+\Delta)^2}
\nonumber \\
&&
+\frac{5\Delta^2-8\Delta m+8m^2}{4g^2\Delta^4 m^4}S_1^2
+\frac{5\Delta^2+8\Delta m+8m^2}{4g^2\Delta^4(m+\Delta)^4}S_2^2
\nonumber \\
&&
+\frac{2\Delta^4+3\Delta^3m-5\Delta^2m^2-16\Delta m^3-8m^4}
{g^2\Delta^4m^3(m+\Delta)^3}S_1S_2
+{\cal O}(S^3).
\end{eqnarray}
These terms precisely coincide with
(\ref{Graph1}), (\ref{Graph2}) and (\ref{Graph3}). 
The absence of the  contribution from $\omega^{(2)}(z)$ can 
also be checked to all orders by numerical calculations.

\section{CFT Techniques}
So far we have considered the planar contributions 
to the $\mbox{Tr}R\wedge R$ corrections in 
supersymmetric gauge theories. 
On the other hand, there are also corrections coming from the 
torus diagrams \cite{Bessis,ACKM,Chekhov,Eynard}, which are known to be elegantly computed using 
CFT techniques \cite{Douglas,Kostov,DST}. In this section 
we will see how the full $\mbox{Tr}R\wedge R$ correction $\F_{\chi=0}$
($=$ the planar 
gravitational corrections $\F^{\mbox{\scriptsize (pl)}}$ (\ref{Fplgr}) 
+ the torus contribution $\F_1^{(0)}$) is reproduced in this framework
for the gauge theories with matter.  
We will first recall how it works in the case without matter \cite{FKN,IM,MM,AJM},
and then examine how the matter fields fit in the story.

\subsection{CFT Techniques without Matter}
Consider the matrix model partition function\begin{eqnarray}
Z&=&\int d \Phi  
\exp \left(-\frac1{g_s} {\rm Tr}W( \Phi ) 
\right)\label{Z}
\end{eqnarray}
with an arbitrary tree level potential 
\begin{eqnarray}W(\Phi)=-\sum_{k=0}^\infty t_k \Phi^k.\end{eqnarray}
CFT techniques are based 
on the the equivalence of the loop equation and the Virasoro constraints
\cite{FKN,IM,MM,AJM}. 
That is, defining the loop operator
\begin{eqnarray}
\omega(z)&:=&\frac1{\hat N} \mbox{Tr}
\frac 1{z-\Phi}
\end{eqnarray}
and the corrective field
\begin{eqnarray}
\varphi(z)&:=& -W(z) +2 g_s\mbox{Tr}\log(z-\Phi),
\\
\partial\varphi(z)&=& 
2S \omega(z)  - W'(z),\nonumber
\end{eqnarray}
the loop equation 
\begin{eqnarray}
\left<
\omega(z)^2
\right>
&=&\frac1{g_s}\sum_{i=1}^{\hat N}\left<
\frac{W'(\lambda_i)}{z-\lambda_i}
\right>
\label{Vconstraint}
\end{eqnarray}
can be written in the equivalent form 
\begin{eqnarray}
\left<\frac1{2\pi i}\oint_C dw \frac{(\partial\varphi(w))^2}{w-z}
\right>&=&0,
\end{eqnarray}
where $<\cdots>$ denotes the expectation value. 
Here the contour $C$ encircles all $w=\lambda_i$, the eigenvalues of $\Phi$, 
but not $w=z$. Since 
\begin{eqnarray}\partial\varphi(z)&=&\sum_{n=0}^{\infty} n t_n z^{n-1} 
+\sum_{n=-\infty}^{-1}2 g_s^2 z^{n-1}\frac{\partial}{\partial t_n}\nonumber\end{eqnarray}
inside a correlator,  $\varphi(z)$ may be regarded as a free chiral boson of 
conformal field theory \cite{Chiral},
and the equation (\ref{Vconstraint}) can be written as the Virasoro 
constraints
\begin{eqnarray}\left<
\oint_C dw\frac{T_{\mbox{\scriptsize em}}(w)}{w-z}
\right>&=&0,
\label{vira}
\end{eqnarray}
where
\begin{eqnarray}T_{\mbox{\scriptsize em}}(z)
&=&+\frac1{4g_s^2}(\partial\varphi(z))^2 
\label{energy-momentum}\end{eqnarray}
is the energy-momentum tensor. Note that the prefactor $\frac1{4g_s^2}$ 
has been chosen so that their moments generate the Virasoro algebra in
the standard normalization.

To leading order in $g_s$, $Z$ is computed by the saddle point 
approximation in which $\partial \varphi(z)$ is replaced with its large $\hat N$
expectation value $-y^{(0)}(z)$ \cite{DV}.
We need to compute the next-to-leading order, which may be obtained by
the Gaussian approximation around a classical solution \cite{Kostov}, 
and hence is described by a {\it free} conformal field theory.

For the case of pure gauge theories without matter, CFT techniques were utilized 
\cite{DST} to compute the genus one contribution to the $\mbox{Tr}R\wedge R$ correction. 
This goes as follows:
Consider the correlation function of $2n$ twist operators 
$\sigma(a_i, \bar a_i)$ 
$(i=1,\ldots,2n)$ on a sphere \cite{DFMS,Z,BR,VV,DVV} 
\begin{eqnarray}
\left<
\prod_{k=1}^{2n}\sigma(a_k,\bar a_k)
\right>^{\rm \!\!CFT}
&=&|\det A|^{-1}\prod_{i<j}^{2n}|a_i-a_j|^{-\frac14}
\sum_{(\{p_i\},\{\bar p_i\})}\exp\left[
\frac12 i\pi(p\cdot\tau\cdot p -\bar p\cdot\bar\tau\cdot\bar p)
\right].\nonumber\\
\label{A}\end{eqnarray}
where $\tau$ is the period matrix of the double cover with genus $n$ 
(not $n\!-\!1$!; the extra handle arises by the plumbing fixture 
connecting $z=\infty$ and $\tilde\infty$. See Section \ref{Remark}.)
and 
\begin{eqnarray}
A_{ij}&:=&\oint_{A_i}dz\frac{z^{j-1}}y.~~~
\end{eqnarray}
The loop momenta $\{p_i\}_{i=1,\ldots,n-1}$ run over a certain 
momentum winding lattice.

It was argued in \cite{Kostov}, and was recently confirmed in \cite{Chekhov}, 
that the genus one free energy can be built 
from the chiral determinant of a $c=1$ free boson CFT, or  
the chiral piece of (\ref{A}) with particular loop momenta  
\begin{eqnarray}\left<
\prod_{k=1}^{2n}\sigma(a_k)
\right>^{\rm \!\!CFT}_{p_i=-\sqrt 2 N_i}
&=&e^{\pi i N\cdot\tau\cdot N}\mbox{det}\Delta_0^{-\frac12} \nonumber 
\\
&=&
e^{\pi i N\cdot\tau\cdot N}\det A^{-\frac12}
\prod_{i<j}^{2n}(a_i-a_j)^{-\frac18}.
\label{<sigma>}\end{eqnarray}
The important point is that the twist operator $\sigma(z)$ does not 
satisfy the Virasoro constraint (because, for example, it has nonzero 
conformal weight $\frac1{16}$) and hence must be replaced with an
appropriately modified `star' operator $S(z)$ \cite{Moore} given by 
\begin{eqnarray}S(z)&=&\mu_{\frac32}(z)^{-\frac1{24}}\sigma(z),
\label{star}
\\
<\partial\varphi(z)>&=&\sum_{r\geq\frac12}
\mu_r(a_i)~(z-a_i)^{r-1}\end{eqnarray}
to this order of the approximation. 
If the number of cuts $n$ is equal to the degree of 
$W'(\Phi)$,  we find\footnote{If the number of cuts $n$ is smaller than 
the degree of $W'(\Phi)$, 
(\ref{logmu}) has additional contributions corresponding to the 
moment factors in eq.(4.9) of \cite{Chekhov}, which reduce to 
constants in the maximally broken case.} 
\begin{eqnarray}
\sum_{i=1}^{2n}\log\mu_{\frac32}(a_i)&=&
\log\prod_{i<j}(a_i-a_j),
\label{logmu}
\end{eqnarray}
The replacement (\ref{star}) then correctly yields the genus one free energy
\begin{eqnarray}e^{\F^{\mbox{\scriptsize (pl)}}+
\F_1^{(0)}}&=&\left<
\prod_{k=1}^{2n}S(a_k)
\right>^{\rm \!\!CFT}_{p_i=-\sqrt 2 N_i}\nonumber\\
&=&
e^{\pi i N\cdot\tau\cdot N}\det A^{-\frac12}
\prod_{i<j}(a_i-a_j)^{-\frac16}
\label{e^F1pure}\end{eqnarray}
as was confirmed in \cite{ACKM,Chekhov}.\footnote{$\mbox{det}A$ reduces to that of genus $n-1$ up to an irrelevant 
multiplicative constant in the $|\Lambda_0|\rightarrow\infty$ limit.}
Note that  the CFT result 
(\ref{e^F1pure}) also automatically includes the planar 
gravitational correction from sphere diagrams 
\cite{DGOVZ} with the momentum identification\footnote{There are reasons why the assignment (\ref{p=sqrt2N})
is natural. First, $T(z):=\left<\mbox{Tr}\frac1{z-\phi}\right>$ 
has a total residue $N_f$ from matter poles \cite{CSW}
associated with the
total $\frac{N_f}{\sqrt{2}}$ momentum of vertex operators 
(See Section \ref{fermion}.), whereas 
the difference of  the residues at $z=\infty$ and $\tilde \infty$ is $2N$.
Another related reason is that if $N_f=2N$, the total momentum vanishes 
at infinity with this assignment, naturally reflecting the conformal 
bound of the gauge theory. }
\begin{eqnarray}
p_i&=&-\sqrt 2 N_i~~~ (i=1,\ldots,n).
\label{p=sqrt2N}
\end{eqnarray}

\subsection{The Full $\mbox{Tr}R\wedge R$ 
Correction and Fermion Correlators}
\label{fermion}
We next consider the CFT interpretation of the gravitational corrections 
for gauge theories with matter.
The partition function is now
\begin{eqnarray}
Z&=&\int d \Phi  
\exp \left(-\frac1{g_s} {\rm Tr}W( \Phi ) \right)
e^{-\sum_{I=1}^{N_f}\mbox{\scriptsize Tr}\log(\Phi -m_I)}.
\label{Zmatter}
\end{eqnarray}
Treating the second factor perturbatively, it is an expectation value 
of $e^{-\sum_{I=1}^{N_f}\mbox{\scriptsize Tr}\log(\Phi -m_I)}$ 
with respect to the measure without matter.
Since
\begin{eqnarray}\mbox{Tr}\log(\Phi-m_I)&=&\frac1{2 g_s}(\varphi(m_I)+W(m_I))
+\mbox{Tr}\log(-1),\end{eqnarray}
it should be obtained as a correlator of vertex operators charge 
$-\frac1{2g_s}$. If we take the convention that
\begin{eqnarray}<\varphi(z)\varphi(w)>^{\mbox{\scriptsize CFT}}&\sim&-\log(z-w),\end{eqnarray}
the prefactor of (\ref{energy-momentum}) must be $-\frac12$ so that 
$g_s^2=-\frac12$. 
This motivates us to consider the CFT correlator
\begin{eqnarray}
\left<
\prod_{k=1}^{2n}\sigma(a_k,\bar a_k)
\prod_{I=1}^{N_f}e^{\frac 1{\sqrt{2}}i\varphi(m_I)}\cdot
e^{-\frac{N_f}{\sqrt 2}i\varphi(\infty)}
\right>^{\rm \!\!CFT}.
\label{<VV>}
\end{eqnarray}
This is equivalent to a correlation function of conjugate pairs of the 
vertex operators on a hyper-elliptic Riemann surface without twist operators, 
except that their zero-mode contributions differ by a factor of two \cite{DVV,Bernard}.  
In fact, to correctly reproduce the annulus amplitude 
we also need to take the normal ordering of the vertex operators on the same sheets :  
\begin{eqnarray}&&\left<
:\prod_{I=1}^{N_f}e^{\frac 1{\sqrt{2}}i\varphi(m_I)}\cdot
e^{-\frac{N_f}{\sqrt 2}i\varphi(\infty)}:
:\prod_{I=1}^{N_f}e^{-\frac 1{\sqrt{2}}i\varphi(\tilde m_I)}\cdot
e^{\frac{N_f}{\sqrt 2}i\varphi(\tilde\infty)}:
\right>^{\rm \!\!CFT}\nonumber\\
&&\quad\quad\quad\quad =(\mbox{det}\Delta_0)^{-1} 
\sum_{
\scriptsize
\begin{array}{c}
(\{p_i\},\{\bar p_i\})
\end{array}}A_0^{\{p_i\}} {\overline A_0^{\{p_i\}}
} , 
\label{CFTcorrelator}\end{eqnarray} 
where $A_0^{\{p_i\}}$ is the holomorphic block with loop momenta 
$\{p_i\}$ $(i=1,\ldots,n)$. 
Again, let us focus on a particular block of (\ref{CFTcorrelator}) with 
$p_i=-\sqrt 2 N_i$  $(i=1,\ldots,n)$.
Taking the logarithm of this block, we have 
\cite{DVV}
\begin{eqnarray}
\log A_0^{\{-\sqrt 2 N_i\}}&=&2\left(
\pi i\sum_{i,j=1}^n\tau_{ij}N_iN_j
+2\pi i \sum_{i=1}^n
N_i \sum_{I=1}^{N_f} 
\int_{m_I}^\infty \hat\omega_i\right)
\nonumber\\
&&
+\frac12\sum_{I,J=1}^{N_f}\log
\frac{E(m_I,\tilde \infty)E(\infty,\tilde m_J)}
{E(m_I,\tilde m_J)E(\infty,\tilde \infty)}.
\label{logA0}
\end{eqnarray}
$\tau_{ij}$ is the period matrix of the Riemann surface on which 
the CFT is defined (See Section \ref{Remark}.).
The first and second terms are twice of the planar gravitational 
corrections (\ref{dF2dSdS})(\ref{NdFdS}). As we mentioned, the zero-mode part of 
the twist correlator is half of that on its double cover 
(eqs.(4.3) and (5.21) in \cite{DVV}).
The last term is equal to
the annulus amplitude $\F_0^{(2)}$ 
(\ref{F0(2)primeform}); 
by definition of normal ordering, it does not contain the 
correlations between pairs of points on the same sheet. 
The `star'ization of the matter vertices is not necessary,
since the contour $C$ in (\ref{vira}) does not encircle $z=m_I$.
Thus
we have shown that the chiral block (\ref{logA0}) together with the 
corrected chiral determinant
precisely reproduces 
all the contributions to the full $\mbox{Tr}R\wedge R$ correction 
for the gauge theories with matter.

 Remarkably, the correlator (\ref{<VV>}) closely resembles that in the 
bosonization of fermions on Riemann surfaces \cite{Bernard}. 
This is in accord with the fact that the correlators of $\det (\Phi-m_I)$ can be 
described as free fermion insertions in the soliton theory \cite{Kyoto}, and in 
some special case they are realized as CFT correlators \cite{IMO}. 
However, a significant difference here
is that they are all for the full matrix model correlators, whereas our
result concerns only the contributions of ${\cal O}(1/\hat N^2)$. Another important 
point is that the normal ordering is required in (\ref{CFTcorrelator}).

\subsection{Non-compact Calabi-Yau as a Pinched Riemann Surface}
\label{Remark}  
The ${\cal N}=1$ supersymmetric 
gauge theories are realized in type IIB superstring theory on non-compact 
Calabi-Yau three-folds.
If  one computes the effective superpotential by using the geometric 
transition,  one needs to introduce a cutoff $\Lambda_0$ 
to regularize some periods of  a  reduced {\it non-compact} 
Riemann surface.  Although the quantities computed 
in matrix models are finite (since the eigenvalues are distributed on finite 
intervals), a cutoff is needed again if the free energy (or the disk amplitude) 
is written as a contour integral around infinity.
Thus a matrix model curve with $n$ cuts is a genus $n-1$ 
{\it punctured} Riemann surface. 

On the other hand, the CFT correlator formula obtained in \cite {DVV} and 
used above assumes that the Riemann surface on which the CFT is defined 
is {\it compact} with no punctures. How can we understand the cutoff $\Lambda_0$ here? In fact, the Riemann surface for the CFT can be 
thought of as a {\it pinched} Riemann surface of genus $n$, 
where 
$|\Lambda_0/\Lambda|^{-2}$ 
is identified as the the pinching parameter $t$ of the plumbing fixture. 
Indeed, suppose that we identify two annular regions near 
$\infty$, $\tilde\infty$
\begin{eqnarray}W_1&=&\left\{  p\in{\it \Sigma}^{g_s=0}_{n-1} \mbox{~on the 1st sheet}~\left|~
|\Lambda|
< |z(p)|<\left|\frac{\Lambda_0^2}
{\Lambda
}\right|
\right.\right\},
\nonumber\\
W_2&=&\left\{  p\in{\it \Sigma}^{g_s=0}_{n-1} \mbox{~on the 2nd sheet}~\left|~
|\Lambda|
<|z(p)|<\left|\frac{\Lambda_0^2}
{\Lambda
}\right|\right.\right\}\end{eqnarray}
by gluing them with a cylinder
\begin{eqnarray}
S&=&\{
(X,Y)\in{\mathbb C}^2
~|~XY=t\}
\end{eqnarray}
through the identification
\begin{eqnarray}(X,Y)&=&\left(
\frac{|\Lambda|
}z, 
\frac{tz}{|\Lambda|
}
\right) \mbox{~~~for $W_1$},\nonumber\\
(X,Y)&=&\left(
\frac{tz}{|\Lambda|
},
\frac{|\Lambda|
}z
\right) \mbox{~~~for $W_2$}\end{eqnarray}
with 
$\sqrt t =|\frac{\Lambda}{\Lambda_0}|$.
Then $\hat B_i$'s 
$(i=1,\ldots,n)$ become {\it closed} contours, and we can verify 
by a change of the canonical homology basis that 
the period matrix 
$\{-
\int_{\hat B_i}dz\; \hat\omega_j
\}_{i,j=1,\ldots,n}$ coincides with Fay's  period matrix formula 
for the pinched Riemann surface \cite{Fay} in the 
$|\Lambda_0|\rightarrow \infty$ limit. 

\section{On-shell Condition}
\subsection{The ${\cal N}=2$ Limit and the Torus Contribution}
The ${\cal N}=2$ supersymmetry can be restored for $U(N)$ gauge theories if the superpotential is in the form 
\begin{eqnarray}
W(\phi)=g_{N+1}\left(
\phi^{N+1}+\sum_{p=1}^{N}\frac{g_p}{p}\phi^p
\right)
\end{eqnarray}
so that the gauge breaking pattern  is 
$U(N)\to U(1)^N$;
an ${\cal N}=2$ theory can be 
realized by taking the 
$g_{N+1}\rightarrow 0$ limit \cite{CV}. 
The Seiberg-Witten curve is recovered from the matrix model curve in
this limit if the moduli $S_i$ satisfy the on-shell condition 
$\partial W_{\rm eff}(S_i)/\partial S_i=0$ \cite{NSW,CSW}. (See also \cite{Matone}.)
Turning off the superpotential term in this way, 
the ${\cal N}=1$ Weyl superfield $G_{\alpha\beta\gamma}(\theta)$ and 
the graviphoton superfield 
$F_{\alpha\beta}(\theta)$ are combined into a single 
${\cal N}=2$ Weyl superfield
$W_{\alpha\beta}(\theta,\psi)=
F_{\alpha\beta}(\theta)+\psi^{\gamma}G_{\alpha\beta\gamma}$.
Consequently, 
the ${\cal N}=1$ gravitational F-terms yield to an ${\cal N}=2$ F-term
\begin{eqnarray}
\int d^4xd^4\theta \F_{\chi=0}({\cal S}_i)
W_{\alpha\beta}W^{\alpha\beta}.
\end{eqnarray}
One can thus recover the ${\cal N}=2$ results by simply  
replacing the matrix model curve with the Seiberg-Witten curve.

The $W^2$ factor contains 
the $\mbox{Tr}R\wedge R$ component; it
can be calculated by  topologically twisted ${\cal N}=2$ 
Yang-Mills theories
\cite{MW},  
since the effect of the topological twist becomes invisible on a 
hyper-K\"ahler four-manifold $M_4$ and
the coefficient of 
$\mbox{Tr}R\wedge R$ term of physical ${\cal N}=2$ theories 
agrees with that of topological theories.  The latter yields
up to a constant factor 
\begin{eqnarray}
&&\F_{\chi=0}=b(S_i)-\frac{2}{3}c(S_i),
\label{DW}\\
&&b(S_i)=-\frac{1}{2}\log \det A_{ij}, \quad 
A_{ij}=\oint_{A_i} dz
\frac{z^{j-1}}{y},
\nonumber \\
&&c(S_i)=-\frac{1}{12}\log \Delta, \quad
\Delta=\prod_{i<j}^{2N}(z_i-z_j)^2,
\nonumber
\end{eqnarray}
where $b(S_i)$ and $c(S_i)$ are defined on the Seiberg-Witten curve
\begin{eqnarray}
y^2=P_N(z)^2-\Lambda^{2N-N_f}\prod_{I=1}^{N_f} (z-m_I)
=\prod_{i=1}^{2N}(z-z_i).
\end{eqnarray}

On the Seiberg-Witten curve, 
the torus free energy which is  obtained by conformal field theory 
techniques exactly coincides with (\ref{DW}).
Recently, it was confirmed that the topological partition function 
coincides with the multi-instanton counting formula \cite{Nekrasov} for 
$g=1$ \cite{EK}.
Thus we find consistency of our formula 
with the ${\cal N}=2$ SYM results.

\subsection{The Origin of the Planar Contributions?}
We have, however, also the planar gravitational corrections,
which have no counterpart in the instanton calculations of 
${\cal N}=2$ theories. 
Where do these terms come from?
A suggestive observation is that 
all the planar terms can be combined into 
a simple integral 
\begin{eqnarray}
\F^{\mbox{\scriptsize (pl)}}&=&
2\pi i\left(\frac{1}{2}\sum_{i,j=1}^n N_iN_j\tau_{ij}
+\sum_{i=1}^n N_i\sum_{I=1}^{N_f}\int_{m_I}^{\infty}\hat{\omega}_i\right)
+\frac{1}{2}\sum_{I,J=1}^{N_f}\log
\frac{E(m_I,\tilde{\infty})E(\infty,\tilde{m}_J)}{E(m_I,\tilde{m}_J)E(\infty,\tilde{\infty})}
\nonumber \\
&=&
-\pi i
\left(\sum_{i=1}^nN_i\int_{\hat{B}_i}
-\sum_{I=1}^{N_f}\int_{m_I}^{\infty}\right)dz\; T(z),
\label{annular integral}
\end{eqnarray}
where
\begin{eqnarray}
T(z)\;dz
&=&\sum_{i=1}^{n}N_i\hat\omega_i+\frac1{2\pi i}\sum_{I=1}^{N_f}
\omega_{\tilde{m_I}-\tilde{\infty}}
\label{T(z)}
\end{eqnarray}
is nothing but the gauge theory expectation value of $\left<
\mbox{Tr}\frac1{z-\phi}
\right>$ and solves the Konishi anomaly equations \cite{CSW}!
Since the B-period integrals of  $T(z)$ on shell may naturally be 
regarded as NS-NS fluxes on a deformed Calabi-Yau three-fold \cite{CIV} 
\begin{eqnarray}
W^{\prime}(z)^2+f_{n-1}(z)+y^2+v^2+w^2=0, \quad 
(z,y,v,w)\in \mathbb{C}^4,
\end{eqnarray}
it suggests that 
$\F^{\mbox{\scriptsize (pl)}}$ would correspond to some 
interaction term on the D5-branes couple to 
2-form field $B_2=B_R+\tau B_{NS}$ before the geometric 
transition.  This motivates us to consider  
the induced Chern-Simons term 
on D$p$-branes  \cite{Green}
\begin{eqnarray}
I=\int_{M_{p+1}}C\wedge {\rm ch}(F)\wedge \sqrt{\hat{A}(R)},
\end{eqnarray}
where $C$ is a form field on the D$p$-branes.
${\rm ch}(F)$ and $\hat{A}(R)$ are defined 
for the Chan-Paton gauge  and  tangent bundles on $M_{p+1}$,
respectively.
In our case D5-branes are wrapped around 2-cycles 
in a Calabi-Yau three-fold and filling the four-dimensional space-time.
From this Chern-Simons term we obtain
the $\mbox{Tr}R\wedge R$ term
\begin{eqnarray}
\int_{M_6}C\wedge ({\rm tr}{\bf 1}) \mbox{Tr}R\wedge R,
\end{eqnarray}
where ${\rm tr}{\bf 1}$ comes from ${\rm ch}(F)$
and $C$ is restricted to a 2-form in the Calabi-Yau direction.
In this brane setup 
we will have the $\mbox{Tr}R\wedge R$ term
\begin{eqnarray}
\left(
\sum_{i=1}^nN_i\int_{\hat B_i}
+\sum_{I=1}^{N_f}\int_{\tilde m_I}^{\tilde\infty}
\right)
B_{\rm NS}
\int_{\mathbb{R}^4}\mbox{Tr}R\wedge R.
\label{R^R}
\end{eqnarray}
Note that the weight factors  $N_i$ naturally arise from 
${\rm tr}{\bf 1}$. 
Thus, we conjecture that the on-shell planar gravitational 
corrections may have their origin in 
the Chern-Simons coupling induced on the D5-branes.

Although these arguments are not conclusive, the relation 
(\ref{annular integral}) is 
certainly suggestive and would be worth being studied.

\section{Summary and Discussion}
In this paper we have studied the gravitational corrections to the effective 
superpotential in four-dimensional ${\cal N}=1$ $U(N)$ gauge theories 
with flavors in the fundamental representation, 
using the matrix model approach of Dijkgraaf and Vafa.
We derived a compact formula for the annulus contribution to 
the corrections in terms of the prime form on the matrix model curve.
We also showed that the full $\mbox{Tr}R\wedge R$ correction containing
the torus as well as all the planar contributions can be reproduced as 
a special momentum sector of a single $c=1$ CFT correlator.
The ${\cal N}=2$ limit of the torus contribution agrees with the answer 
of the multi-instanton calculations 
and also with the geometric engineering argument from the topological A-model.
The planar contributions, on the other hand, have no counterpart 
in the instanton calculations of 
${\cal N}=2$ gauge theories, and we speculated that the latter might 
correspond to the Chern-Simons term induced on the D5-branes.

The CFT correlator we found is very close to the fermion 
correlator on the Riemann surface.  
In \cite{integrable} the correspondence between the
non-compact B-branes and fermions has been discussed.
Since the matter contributions come from the $N_f$ D5-branes 
wrapped around non-compact 2-cycles,
our result is consistent with the picture \cite{integrable,Sakai}.
In the topological B-model, 
the genus one partition function
is given by a generalized holomorphic Ray-Singer torsion 
on the Calabi-Yau geometry \cite{BCOV}
\begin{eqnarray}
{\cal F}_1=\sum_{p,q=1}^3pq(-1)^{p+q}\log \det\Delta_{p,q},
\end{eqnarray}
where $\Delta_{p,q}$ is the Laplacian acting on $(p,q)$-forms.
By using Quillen's anomaly, we also find the one-loop open 
topological string partition function.
Our formula may lead to the explicit expression
for it on a non-compact Calabi-Yau three-fold.

There are various interesting directions to extend our analysis 
carried out in this paper. 
A simple extension of our analysis will be to investigate 
the gravitational corrections to the  $SO/Sp$ gauge theories. 
If the ${\cal N}=2$ theory is deformed by an adjoint chiral 
superfield, the Chern-Simons coupling term is also induced 
by the orientifold plane \cite{Mukhi}.
It  would be interesting to see whether such terms arise from  
the Klein bottle amplitudes in orthogonal matrix models.
On the other hand, real symmetric/symplectic matrix models describe 
non-perturbative aspects of the $SO/Sp$ gauge theories with flavors 
in the symmetric/antisymmetric tensor representations, respectively. 
The CFT description of these matrix models can be given by a 
$c=-2$ CFT \cite{Odake}. We will report on this elsewhere.

Another interesting issue to consider concerns 
higher dimensional gauge theories.
It was proposed in \cite{DV4,Wijnholt} that some five dimensional 
gauge theories can be described by unitary matrix models.
The associated matrix model curve 
is represented by a pair of cylinders,
where $z=+\infty$ and $-\infty$ cannot be distinguished.
As a result there exist two chiral boson fields in the CFT 
description.
It is interesting to investigate whether the five dimensional 
multi-instanton calculations can be reproduced in this
CFT analysis. 
We also expect that 
the meaning of the five dimensional Chern-Simons 
term \cite{Tachikawa,Wijnholt}
will be clarified.

\vspace{1cm}

\noindent{\large {\bf Acknowledgements:}}
We would like to thank M. Naka
for his collaboration at an early stage of this work,
and N. Ishibashi and H. Kanno for useful discussions. 
We also thank
H. Eguchi, T. Eguchi, 
S. Hosono,
S. Iso,
K. Ito,
M. Jinzenji,
A. Kato,
H. Kawai, T. Kawai,
Y. Kitazawa,
Y. Matsuo,
Y. Nakayama,
M. Natsuume,
Y. Ookouchi,
H. Suzuki,
Y. Sugawara,
Y. Tachikawa,
Y. Yamada and
A. Yamaguchi 
for helpful comments.
The research of S.M. was supported in part by 
Grant-in-Aid for Scientific Research (C)(2) \#14540286 
from Ministry of Education, Culture, Sports, Science and 
Technology.


\section*{Appendix A ~}\setcounter{equation}{0}
\def\theequation{A.\arabic{equation}}
\def\thesection{A}
In this Appendix we give an alternative proof of (\ref{F02integral}). 
See Section \ref{largeN} for the definitions of $\omega^{(2)}(z)$, $y^{(2)}(z)$, etc.
\subsection{Singularities of $\omega^{(2)}(z)dz$}
\label{subsec:omega2}
We will first determine $\omega^{(2)}(z)dz$ 
from its singularities. In fact, to prove the assertion, 
the precise form is not needed. We will nevertheless derive it for 
future convenience for it is relevant to the computation 
of $\F_0^{(3)}$.

First of all, $\omega^{(2)}(z)dz$ (as well as other higher $\omega^{(k)}(z)dz$'s 
for all $k\geq 1$) has zero A-periods, for the same reason as before ------ 
$S_i$'s are independent of $g_s$. 
In fact, $\omega^{(2)}(z)dz$ has only singularities at the $2n$ branch points 
$z=a_i(0)$ $(i=1,\ldots,2n)$, which are of second order. 
They arise since the matter 
insertions change the locations of the cuts.
We will show that 
$\omega^{(2)}(z)dz$ is an abelian differential of the second kind 
with zero A-periods and given by
\begin{eqnarray}\omega^{(2)}(z)
&=&\frac1{\sqrt{\prod_{l=1}^{2n}(z-a_l)}}
\sum_{i=1}^{2n} \frac{{a'_i}^2M_i^{(1)}
\prod_{\scriptsize 
\begin{array}{c}k\!\!=\!\!1\\ k\!\!\neq\!\! i
\end{array}}^{2n}(a_i - a_k)}{16S(z-a_i)}
+\frac{\sum_{j=0}^{n-2}p_jz^j}{\sqrt{\prod_{l=1}^{2n}(z-a_l)}}
\label{omega(2)}\end{eqnarray}for some $p_j
$ $(j=0,\ldots,n-2)$, 
where $a_i:= a_i(0)$, $a'_i:= a'_i(0)$ and
\begin{eqnarray}
M_i^{(1)}&=&\frac1{2\pi i}\oint_{A_\infty}dz\frac{W'(z)}{(z-a_i)
\sqrt{\prod_{l=1}^{2n}(z-a_l)}}~~~(i=1,\ldots,2n).
\end{eqnarray}
$M_i^{(1)}$'s are the {\it moments} introduced in 
\cite{ACKM}. $p_j$'s $(j=0,\ldots,n-2)$ are determined 
so that $\omega^{(2)}(z)$ has zero A-periods.

Instead of studying the singularities of $\omega(z,g_s)$
directly, it is easier to investigate those of $y(z,g_s)$ 
(\ref{yzgs}). 
$y^{(2)}(z)$ and $-2S\omega^{(2)}(z)$ have the same singularities 
since the difference between the two 
is only linear in $g_s$.

It is easy to see that $y^{(2)}(z)dz$ is regular 
at $z=m_I$, $\tilde m_I$ $(I=1,\ldots,N_f)$ since
$\sum_I\frac{g_s}{z-m_I}$ has no singularity of ${\cal O}(g_s^2)$.
The integral representation  (\ref{yzgs}) shows that $y^{(2)}(z)dz$ has
possible singularities at $z=\infty$, $\tilde\infty$ 
and the branch points $z=a_j(0)$
($j=1,\ldots,2n$). However, using the conditions (\ref{1/zconditions}) as 
well as those obtained by acting $\frac\partial{\partial g_s}|_{g_s=0}$ and  
$\frac{\partial^2}{\partial g_s^2}|_{g_s=0}$, it can be shown that the 
singularities at $z=\infty$, $\tilde\infty$ precisely cancel. 
Therefore, we have only to examine the 
singular behavior of $y^{(2)}(z)$ at the branch points.

Since 
\begin{eqnarray}y^{(2)}(z)&=&\frac12\left.\frac{\partial^2 y(z,g_s)}
{\partial g_s^2}\right|_{g_s=0},\end{eqnarray}we see that the singularities at the branch points arise when  
$\left.\frac{\partial^2}{\partial g_s^2}\right|_{g_s=0}$ acts on 
the first factor $\sqrt{\prod_{i=1}^{2n}(z-a_i(g_s))}$ of $y(z,g_s)$,
and therefore $y^{(2)}(z)$ is proportional to
\begin{eqnarray}&&\frac1{2\pi i}
\left(
\oint_{C}+\oint_{C_z}
\right)
dw \left.\frac{W'(w)
+\sum_{I=1}^{N_f}\frac{g_s}{w-m_I}}
{(z-w)\sqrt{\prod_{i=1}^{2n}(w-a_i(g_s))}}\right|_{g_s=0}
\nonumber\\
&&~~~~~~~~~~~~=
\frac1{2\pi i}
\oint_{A_\infty} dw\frac{W'(w)}
{(z-w)\sqrt{\prod_{i=1}^{2n}(w-a_i(0))}}
\nonumber\\
&&~~~~~~~~~~~\stackrel{z\rightarrow a_j}{\longrightarrow}~
-M_j^{(1)}~~~(j=1,\ldots,2n).
\end{eqnarray}
The contour $C$ surrounds all the cuts, and $C_z$
surrounds $w=z$.
Matching the Laurent coefficients at the branch points, we obtain
 \begin{eqnarray}
 y^{(2)}(z)
&=&-\frac1{\sqrt{\prod_{l=1}^{2n}(z-a_l)}}
\sum_{i=1}^{2n} \frac{{a'_i}^2M_i^{(1)}}{8(z-a_i)}
\prod_{\scriptsize\begin{array}{c}k\!\!=\!\!1\\ k\!\!\neq\!\! i
\end{array}}^{2n}(a_i - a_k)
+\mbox{holomorphic differentials}.\nonumber\\
\end{eqnarray}
This implies the equation (\ref{omega(2)}).

Unlike $\omega^{(1)}(z)$, $\omega^{(2)}(z)$ depends on the 
potential $W(\Phi)$ not only through the positions of the cut 
but also through the moments $M_i^{(1)}$, reflecting the 
non-universality of ${\cal F}^{(3)}_0$.

\subsection{An Alternative Proof of (\ref{F02integral})}
We will now prove (\ref{F02integral}). 
Using the saddle point equation
\begin{eqnarray}
\frac{\partial W(\lambda, g_s)}{\partial \lambda}&=&
2S\int_{-\infty}^\infty d\lambda'\frac{\rho(\lambda',g_s)}{\lambda-\lambda'},
\label{saddlept}
\end{eqnarray}
the large $\hat N$ free energy (\ref{F0rhorholog}) is written as 
\begin{eqnarray}
\F_0(g_s)=\sum_{j=1}^n
S_j\left(
\int_{-\infty}^\infty d\lambda \rho(\lambda,g_s) 
\left(-\frac12 W(\lambda, g_s)+S\log\frac{|\lambda-\lambda_{0j}|}\Lambda
\right)
-\frac12 W(\lambda_{0j},g_s)\right).
\label{F0(gs)}
\end{eqnarray} 
Note that (\ref{saddlept}) is true only if $\lambda$ belongs to the 
nonzero support of $\rho(\lambda, g_s)$, that is,  
only if $\lambda$ is such that 
$\rho(\lambda,g_s)\neq 0$.
$\lambda_{0j}$ is an arbitrary point on the $j$-th cut, whose location 
is independent of $g_s$.

We can extract the annulus contribution 
\begin{eqnarray}
\F_0^{(2)}&=&\F_0^{(2)}({\bf I})+\F_0^{(2)}({\bf II}),\\
\F_0^{(2)}({\bf I})&:=&
-\frac{S}{2}
\int_{-\infty}^\infty d\lambda \rho^{(1)}(\lambda) 
\sum_{I=1}^{N_f}\log(\lambda-m_I),
\nonumber\\
\F_0^{(2)}({\bf II})&:=&-\sum_{j=1}^n\frac{S_j}{2}
\int_{-\infty}^\infty d\lambda \rho^{(2)}(\lambda) 
\left(W(\lambda)-2S\log|\lambda-\lambda_{0j}|\rule{0mm}{4mm}
\right).
\label{F02II}
\end{eqnarray}
($\log\Lambda$ drops out since $\int_{-\infty}^\infty 
d\lambda \rho^{(2)}(\lambda)=0$.)
Since $\F_0^{(2)}({\bf I})$ is already the right hand side of (\ref{F02integral}), 
we must show that $\F_0^{(2)}({\bf II})=0$.
For this purpose 
let us first rewrite (\ref{F02II}) in a contour integral on the first sheet 
using $\omega^{(2)}(z)$. We take the branch cut of $\log(z-\lambda_{0j})$ 
so that it runs from $z=\lambda+i0$ ($z=\lambda$ on the upper side of the 
cut)to $z=-\infty$.
Then depending on $i<j$, $=j$ or $>j$
we have 
\begin{eqnarray}
\hspace*{-0.5cm}
&&\int_{a_{2i-1}}^{a_{2i}} d\lambda\rho^{(2)}(\lambda) 
\log|\lambda-\lambda_{0j}|\nonumber\\
&&\hspace*{-0.5cm}=\left\{
\begin{array}{ll}
{\displaystyle \frac1{2\pi i}\int_{A_i} dz\; \omega^{(2)}(z) 
\left(\log(z-\lambda_{0j})
+\pi i\right)}&(i<j),\\
{\displaystyle \frac1{2\pi i}\left(\int_{A_j(\lambda_{0j})} dz\;\omega^{(2)}(z) 
\log(z-\lambda_{0j})
+\pi i\int_{\lambda_{0j}+i0}^{\lambda_{0j}-i0} dz\; \omega^{(2)}(z) 
\right)}&(i=j),\\
{\displaystyle \frac1{2\pi i}\int_{A_i} dz\;\omega^{(2)}(z) 
\log(z-\lambda_{0j})
}&(i>j),
\end{array}\right.
\label{intrholog}
\end{eqnarray}
where the contour $A_j(\lambda_{0j})$ 
starts from $z=\lambda_{0j}+i0$, surrounds the 
$j$th cut once anti-clockwise and goes back to $z=\lambda_{0j}+i0$ again 
but on the other side of the branch cut of  $\log(z-\lambda_{0j})$.
From (\ref{intrholog}) we find
\begin{eqnarray}
\int_{-\infty}^\infty d\lambda \rho^{(2)}(\lambda) 
\log|\lambda-\lambda_{0j}|
&=&\frac1{4S}\lim_{|\Lambda_0|\rightarrow\infty}
\int_{\hat B_j}
dz\; y^{(2)}(z)~~~(j=1,\ldots,n).
\end{eqnarray}
Therefore, to prove $F_0^{(2)}({\bf II})=0$ it suffices to show 
that 
\begin{eqnarray}
\frac1{2\pi i}\sum_{j=1}^n\oint_{A_j}\!\!
dz\; y^{(2)}(z) W(z) +\sum_{j=1}^n
S_j\!\!\oint_{B_j} 
dz\; y^{(2)}(z)&=&0.
\label{F0(2)reformulated}\end{eqnarray}
This can be proved by applying Riemann's bilinear identity, as we
will show below.

Cut out along
\begin{eqnarray}A_{n-1} B_{n-1} A_{n-1}^{-1} B_{n-1}^{-1}
\cdots A_1 B_1 A_1^{-1} B_1^{-1},
\end{eqnarray}
${\it \Sigma}^{g_s}_{n-1}$ is represented by a 
$4(n-1)$-sided polygon with identification in a standard manner. 
Since $y^{(2)}(z)dz$ is an abelian differential of the second kind, 
its integral 
\begin{eqnarray}
Y(z):=\int_{z_0}^z dx\; y^{(2)}(x)
\end{eqnarray}
defines a single-valued meromorphic function $Y(z)$ inside the polygon 
for an arbitrary reference point $z_0$. Let us evaluate the integral of 
\begin{eqnarray}
\eta:= Y(z)y^{(0)}(z)dz
\end{eqnarray} 
along the $4(n-1)$ sides of the polygon. Applying  
Riemann's bilinear identity, we obtain
\begin{eqnarray}
2\pi i\sum_{p\in{\it \Sigma}^{g_s}_{n-1}}\mbox{Res}_p \eta&=&
\sum_{j=1}^{n-1}\left(\oint_{A_j}\!\!dz y^{(2)}(z)\oint_{B_j}\!\!dx y^{(0)}(x)
-\oint_{B_j}\!\!dz y^{(2)}(z)\oint_{A_j}\!\!dx y^{(0)}(x)\right).\nonumber\\
\label{identity}
\end{eqnarray}
The right hand side is
\begin{eqnarray}\mbox{right hand side of  (\ref{identity})}
&=&4\pi i \sum_{j=1}^{n-1}S_j\oint_{B_j}\!\!dz\; y^{(2)}(z).
\label{rhs}
\end{eqnarray}

Let us compute the left hand side.
Since
\begin{eqnarray}
y^{(0)}(z)&\stackrel{z\rightarrow\infty}{\rightarrow}&
-\frac{2S}z+W'(z)+{\cal O}(z^{-2}),\nonumber\\
&\stackrel{z\rightarrow\tilde\infty}{\rightarrow}&
+\frac{2S}z-W'(z)+{\cal O}(z^{-2}),
\end{eqnarray}  
we find
\begin{eqnarray}
2\pi i\mbox{Res}_\infty \eta&=&
\oint_{A_\infty} dz\; y^{(2)}(z)W(z)+4\pi i S Y(\infty),\nonumber\\
2\pi i\mbox{Res}_{\tilde\infty} \eta&=&
-\oint_{A_{\tilde\infty}} dz\; y^{(2)}(z)W(z)-4\pi i S Y(\tilde\infty)\nonumber\\
&=&
+\oint_{A_\infty} dz\; y^{(2)}(z)W(z)-4\pi i S Y(\tilde\infty),
\end{eqnarray}
where $A_{\tilde\infty}$ is the same as $A_\infty$ but on the second sheet.
We have shown in section \ref{subsec:omega2}
that $y^{(2)}(z)dz$ has second order poles at every branch point
$z=a_i$ $(i=1,\ldots,2n)$, and therefore $Y(z)$ has first order poles 
there, while we also see from (\ref{yzgs}) that $y^{(0)}dz$ has 
first order zeroes at those branch points. Therefore $\eta$ is regular 
there and hence has no nonzero residues. Thus, in all, we have
\begin{eqnarray}
&&\hspace*{-1cm}
\mbox{left hand side of (\ref{identity})}
=2\oint_{A_\infty}dz\; y^{(2)}(z) W(z)+4\pi i S\left(
Y(\infty)-Y(\tilde\infty)\right).
\label{lhs}
\end{eqnarray}
By definition
\begin{eqnarray}Y(\infty)-Y(\tilde\infty)&=&
\int^\infty_{\tilde\infty}dz\; y^{(2)}(z),\end{eqnarray}
where the contour cannot cross any side of the polygonal 
region since $Y(z)$ is single-valued only inside. 
Therefore the contour can pass only through the $n$th cut:
\begin{eqnarray}
Y(\infty)-Y(\tilde\infty)&=&\lim_{|\Lambda_0|\rightarrow\infty}\int_{\hat B_n}dz\; y^{(2)}(z).
\label{Y-Y}
\end{eqnarray}Plugging (\ref{Y-Y}) into (\ref{lhs}) and equating it with (\ref{rhs}), 
we obtain the desired equation
(\ref{F0(2)reformulated}).

\section*{Appendix B ~The Special Geometry 
Relation}
\setcounter{equation}{0}
\def\theequation{B.\arabic{equation}}
It is now well-known that the sphere amplitude $\F_0^{(0)}$ 
satisfies the special geometry relation
\begin{eqnarray}
\frac{\partial \F_0^{(0)}}{\partial S_i}&=&
\lim_{|\Lambda_0|\rightarrow\infty}\left(
\frac12\int_{\hat B_i}dz\; y^{(0)}(z)
-W(\Lambda_0)+ 2S \log \frac{-\Lambda_0}{\Lambda}
\right).
\label{specialgeometry}
\end{eqnarray}
This fact was pointed out and proven using the energy cost argument 
in \cite{DV}. A proof using a Legendre transformation was given 
in \cite{CSW}. In this Appendix we will give a sketch of an alternative 
proof neither using Legendre transformations nor resorting to any
physical argument.

Before we consider (\ref{specialgeometry}), we first recall  the 
relation \cite{Ferrari}
\begin{eqnarray}\frac{\partial y^{(0)}(z)}{\partial S_i}dz&=&-4\pi i \hat \omega_i(z)
~~~(i=1,\ldots,n),  
\label{dy0dS'}
\end{eqnarray}
where, as we defined in the text, 
$y^{(0)}(z)$ is related to the large $\hat N$ resolvent 
$\omega^{(0)}(z)$ by
\begin{eqnarray}
y^{(0)}(z)&=&W'(z)-2S\omega^{(0)}(z).\end{eqnarray}
Note that the relation (\ref{dy0dS'}) is also true when 
the degree of $W'(z)$ exceeds the number of the cut $n$ and 
$y^{(0)}(z)$ has extra zero factors (See footnote \ref{footnote1}.).
To see this we first use the integral representation of $y^{(0)}(z)$ 
(obtained by setting $g_s=0$ in (\ref{yzgs})) to verify that 
the possible singularities of $\frac{\partial y^{(0)}(z)}{\partial S_i}$ 
are only located at $z=\infty$ and $\tilde\infty$. Next we write
\begin{eqnarray}y^{(0)}(z)&=\sqrt{W'(z)^2+f(z,\{S_j\})}\end{eqnarray}
for some $f(z,\{S_j\})$ to find that 
\begin{eqnarray}\frac{f(z,\{S_j\})}{W'(z)}&\stackrel{z\rightarrow\infty}
{\rightarrow}&{\cal O}(\frac1 z)\end{eqnarray}
since
\begin{eqnarray}2S\omega^{(0)}(z)&=&W'(z)-\sqrt{W'(z)^2+f(z,\{S_j\})}\nonumber\\
&=&\frac{-f(z,\{S_j\})}{W'(z)+\sqrt{W'(z)^2+f(z,\{S_j\})}}
\end{eqnarray}
must behave like $\frac{2S}z +{\cal O}(\frac1{z^2})$ as  
$z\rightarrow\infty$. Therefore we find that
\begin{eqnarray}
\frac{\partial y^{(0)}}{\partial S_i}dz&=&
\frac{\frac\partial{\partial S_i} f(z,\{S_j\})}
{2\sqrt{W'(z)^2+f(z,\{S_j\})}}dz
\end{eqnarray}
has at most a first order pole at $z=\infty$, and hence 
can be written as a linear combination of $\hat\omega_i$'s.
Then (\ref{dy0dS'})  follows from the fact that 
the $A_i$-period of $y^{(0)}(z)dz$ is $-4\pi i S_i$ $(i=1,\ldots,n)$.

We will now prove the special geometry relation (\ref{specialgeometry}).
$\F_0^{(0)}$ is given by 
\begin{eqnarray}\F_0^{(0)}&=&
-\frac{S}{2}
\int_{-\infty}^\infty d\lambda \rho^{(0)}(\lambda) 
W(\lambda)
\\
&&
+S\sum_{j=1}^n S_j
\left(
\int_{-\infty}^\infty d\lambda \rho^{(0)}(\lambda) 
\log\frac{|\lambda-\lambda_{0j}|}{\Lambda}
-\frac1{2S} W(\lambda_{0j})\right).
\end{eqnarray}
We express it in terms of contour integrals of the complex plane. 
Note that the integral of 
$dz y^{(0)}(z)\log(z-\lambda_{0j})$ along $A_\infty$ does not 
vanish as $|\Lambda_0|\rightarrow\infty$, and hence must be taken 
into account:
\begin{eqnarray}
\F_0^{(0)}&=&\lim_{|\Lambda_0|\rightarrow\infty}\left[\rule{0mm}{7mm}
\frac1{8\pi i}\oint_{A_\infty}dz\; y^{(0)}(z)W(z)
\right.\nonumber\\
&&~~~~~~\left.
+\sum_{j=1}^n \frac{S_j}4\int_{\hat B_j}dz\; y^{(0)}(z)
-\frac S2 W(\Lambda_0) +S^2\log\frac{-\Lambda_0}{\Lambda}
\right].\end{eqnarray}
Differentiating it by $S_i$, we find 
\begin{eqnarray}\frac{\partial \F_0^{(0)}}{\partial S_j}
&=&\lim_{|\Lambda_0|\rightarrow\infty}
\left[
-\frac12\oint_{A_\infty}
\hat\omega_i(z) W(z)
-\pi i\sum_{j=1}^nS_j\int_{\hat B_j} \hat\omega_i(z)
+S\log\frac{-\Lambda_0}{\Lambda}
\right.\nonumber\\
&&~~~~~~~\left.+\frac14
\int_{\hat B_j}dz\; y^{(0)}(z)
-\frac12 W(\Lambda_0) +S\log \frac{-\Lambda_0}{\Lambda}
\right],\label{dF0Si}\end{eqnarray}
where the $S\log\Lambda_0$ terms are separated so that each line 
gives a finite result. Again, we can simplify it by using Riemann's 
bilinear relation. In this case we cut out the Riemann surface 
along 
\begin{eqnarray}A_{n-1} B_{n-1} A_{n-1}^{-1} B_{n-1}^{-1}
\cdots A_1 B_1 A_1^{-1} B_1^{-1}
A_{\tilde\infty} \hat B_n A_\infty 
(\hat B_n)^{-1}
\label{boundary}\end{eqnarray}
and evaluate
\begin{eqnarray}\int dz\; y^{(0)}(z)\Omega_i(z),~~~
\Omega_i(z):=\int_{z_0}^z
\omega_i(x)~~~(i=1,\ldots,n),~~~\omega_n(z):=\hat\omega_n(z)\end{eqnarray}
along the boundary of the resulting polygon.
The result we obtain is
\begin{eqnarray}
\lim_{|\Lambda_0|\rightarrow\infty}
\left[
\oint_{A_\infty}
\hat\omega_i(z) W(z)
+2\pi i\sum_{j=1}^nS_j\int_{\hat B_j} \hat\omega_i(z)
+\frac12
\int_{\hat B_i}dz\; y^{(0)}(z)
- W(\Lambda_0)
\right]=0 \label{some_relation}
\end{eqnarray}
for $i=1,\ldots,n$. Using (\ref{some_relation}) in (\ref{dF0Si}), 
we obtain the special geometry relation (\ref{specialgeometry}).

\section*{Appendix C ~The Large $\hat N$ Free Energy for a Quadratic Potential}
\setcounter{equation}{0}
\def\theequation{C.\arabic{equation}}

\begin{eqnarray}\F_0&=&\frac S2\left[\frac{t^2\mu(g_s)^2}S\left(
\frac{c(g_s)^2}2 +\frac{\mu(g_s)^2}8
\right)\right.\nonumber\\
&&-\frac{\frac{g_s t}S}{\alpha(g_s)-\alpha(g_s)^{-1}}\left(
\frac{\mu(g_s)(m+c(g_s))}2-\frac{m^2}{\alpha(g_s)}
\right)\nonumber\\
&&
-\frac{g_s t \mu(g_s)^2}{2S}\left(
\log\left(
-\frac{\mu(g_s)\alpha(g_s)}2
\right)+\frac1{2\alpha(g_s)^2}
\right)\nonumber\\
&&-\frac{g_s^2}S\left(
\frac{\alpha(g_s)^{-1}(\log(-\frac{\mu(g_s)}2)-1)+\alpha(g_s)\log\alpha(g_s)}
{\alpha(g_s)-\alpha(g_s)^{-1}}
-\log(\alpha(g_s)-\alpha(g_s)^{-1})
\right)\nonumber\\
&&
+t\mu(g_s)^2\left(
\log\frac{\mu(g_s)}2+\frac{c(g_s)^2}{\mu(g_s)^2}-\frac12
\right)
\nonumber\\
&&
+2g_s\left(
\frac{\frac{m-c(g_s)}{\mu(g_s)}\log\frac{\mu(g_s)}2
+\frac{c(g_s)}{\mu(g_s)}}{\alpha(g_s)-\alpha(g_s)^{-1}}
-\frac12
\log\frac m{\alpha(g_s)}
\right)\nonumber\\
&&\left.\rule{0mm}{6mm}+g_s\log(-m)\right],\end{eqnarray}where     
\begin{eqnarray}\alpha(g_s)^{\pm1}&=&\frac{m-c(g_s)\pm\sqrt{(m-c(g_s))^2-\mu(g_s)^2}}
{\mu(g_s)}.\end{eqnarray}

\end{document}